\documentclass{aa}  

\usepackage{graphicx}
\usepackage{txfonts}
\usepackage{hyperref}
\newcommand{\gaia}{\textit{Gaia}}

\providecommand{\grp}{\ensuremath{G_\mathrm{RP}}}
\providecommand{\gbp}{\ensuremath{G_\mathrm{BP}}}

\begin{document}

   \title{Cepheid Metallicity in the Leavitt Law (C- MetaLL) survey: IV. The metallicity dependence of Cepheid period--luminosity relations}

\titlerunning{C-MetaLL Survey IV.}

   \author{E. Trentin \inst{1,2,3}\thanks{E-mail: etrentin@aip.de}
   \and
   V. Ripepi\inst{3}
    \and
    R. Molinaro \inst{3}
    \and
    G. Catanzaro \inst{4}
    \and 
    J. Storm \inst{1}
    \and
    G. De Somma \inst{3,5}
    \and
    M. Marconi \inst{3} 
    \and
    A. Bhardwaj\inst{3} 
    \and
    M. Gatto\inst{3} 
    \and
    V. Testa\inst{6} 
    \and     
    I. Musella \inst{1}
    \and
    G. Clementini \inst{7}
    \and
    S. Leccia \inst{3}
       }

\institute{Leibniz-Institut f\"ur Astrophysik Potsdam (AIP), An der Sternwarte 16, D-14482 Potsdam, Germany
\and
Institut für Physik und Astronomie, Universität Potsdam, Haus 28, Karl-Liebknecht-Str. 24/25, D-14476 Golm (Potsdam), Germany
\and
INAF-Osservatorio Astronomico di Capodimonte, Salita Moiariello 16, 80131, Naples, Italy
\and
INAF-Osservatorio Astrofisico di Catania, Via S.Sofia 78, 95123, Catania, Italy 
\and
Istituto Nazionale di Fisica Nucleare (INFN)-Sez. di Napoli, Via Cinthia, 80126 Napoli, Italy
\and
INAF – Osservatorio Astronomico di Roma, via Frascati 33, I-00078 Monte Porzio Catone, Italy 
\and
INAF-Osservatorio di Astrofisica e Scienza dello Spazio, Via Gobetti 93/3, I-40129 Bologna, Italy 
             }

   \date{Received September 15, 1996; accepted March 16, 1997}

 
  \abstract
   {Classical Cepheids (DCEPs) play a fundamental role in the calibration of the extragalactic distance ladder, which eventually leads to the determination of the Hubble constant($H_0$) thanks to the period--luminosity ($PL$) and period--Wesenheit ($PW$) relations exhibited by these pulsating variables. Therefore, it is of great importance to establish the dependence of $PL$ and $PW$ relations on metallicity.}
   {We aim to quantify the metallicity dependence of the $PL$ and $PW$ relations of  the Galactic DCEPs for a variety of photometric bands, ranging from optical to near-infrared.}
   {We gathered a literature sample of 910 DCEPs with available [Fe/H] values from high-resolution spectroscopy or metallicities from the \gaia\ Radial Velocity Spectrometer (RVS). For all these stars, we collected photometry in the $G_{BP},G_{RP},G,I,V,J,H$, and $K_S$ bands and astrometry from \gaia\ Data Release 3 (DR3). We used these data to investigate the metal dependence of both the intercepts and slopes of a variety of $PL$ and $PW$ relations at multiple wavelengths.}
    {We find a large negative metallicity effect on the intercept ($\gamma$ coefficient) of all the $PL$ and $PW$ relations investigated in this work, while present data still do not allow us to draw firm conclusions regarding the metal dependence of the slope ($\delta$ coefficient). The typical values of $\gamma$ are around $-0.4:-0.5$ mag/dex, which is larger than most of the recent determinations present in the literature. We carried out several tests, which confirm the robustness of our results. As in our previous works, we find that the inclusion of a global zero point offset of \gaia\ parallaxes provides smaller values of $\gamma$ (in an absolute sense). However, the assumption of the geometric distance of the Large Magellanic Cloud (LMC) seems to indicate that larger values of $\gamma$ (in an absolute sense) would be preferred.}
   {}

   \maketitle
%

\section{Introduction}

 Classical Cepheids (DCEPs) are the most important standard candles of the extragalactic distance scale thanks to the Leavitt Law \citep{Leavitt1912}, which is a relationship between period and luminosity (PL) of DCEPs. Once calibrated using independent distances based on geometric methods such as trigonometric parallaxes, eclipsing binaries, and water masers, these relations constitute the first step in forming the cosmic distance scale, as they calibrate secondary distance indicators. These latter include Type Ia supernovae (SNIa), which in turn allow us to measure the distances of distant galaxies located in the steady Hubble flow.  The calibration of this three-step procedure (also called the cosmic distance ladder) allows us to reach the Hubble flow, where the constant (the Hubble constant $H_0$) that connects the distance to the recession velocity of galaxies can be estimated \citep[e.g.][and references therein]{Sandage2006,Freedman2012,Riess2016}.
The value of $H_0$ is an important quantity in cosmology because it sets the dimension and the age of the Universe. Therefore, measuring the value of the constant with an accuracy of 1\% is one of the most important quests of modern astrophysics. 
However, there is currently a well-known discrepancy between the values of $H_0$ obtained by the SH0ES\footnote{Supernovae, HO, for the Equation of State of Dark energy} project through the cosmic distance ladder \citep[$H_0=$73.01$\pm$0.99 km s$^{-1}$ Mpc$^{-1}$,][]{Riess2022a} and those measured by the Planck Cosmic Microwave Background (CMB) project using the flat $\Lambda$ Cold Dark Matter ($\Lambda$CDM) model \citep[$H_0=$67.4$\pm$0.5 km s$^{-1}$ Mpc$^{-1}$,][]{Planck2020}. No solution has yet been proposed for this 5$\sigma$ discrepancy, and if confirmed, it would highlight the need for a revision of the $\Lambda$CDM model. 

In this context, one of the residual sources of uncertainty in the cosmic distance ladder is represented by the debated metallicity dependence of the DCEP $PL$ relations used to calibrate the secondary distance indicators. Indeed, a metallicity variation is predicted to affect the shape and width of the DCEP instability strip \citep[e.g.][]{Caputo2000}, which in turn affects the coefficient of the $PL$ relations \citep[][and references therein]{Marconi2005,Marconi2010,DeSomma2022}. The dependence of the $PL$ relations and of the reddening-free Wesenheit magnitudes\footnote{The Wesenheit magnitudes introduced by \citet{Madore1982}  provide a reddening-free magnitude by supposing that the adopted extinction law does not change significantly from star to star as can happen if the targets are placed in regions of the Galaxy with different chemical enrichment histories.}  on metallicity, however small, when involving near-infrared \citep[NIR, see e.g.][]{Fiorentino2013,Gieren2018} 
colours must be taken into account to avoid systematic effects in the calibration of the extragalactic distance scale \citep[e.g.][]{Romaniello2008,Bono2010,Riess2016}. 
  \begin{figure}
   \includegraphics[width=9cm]{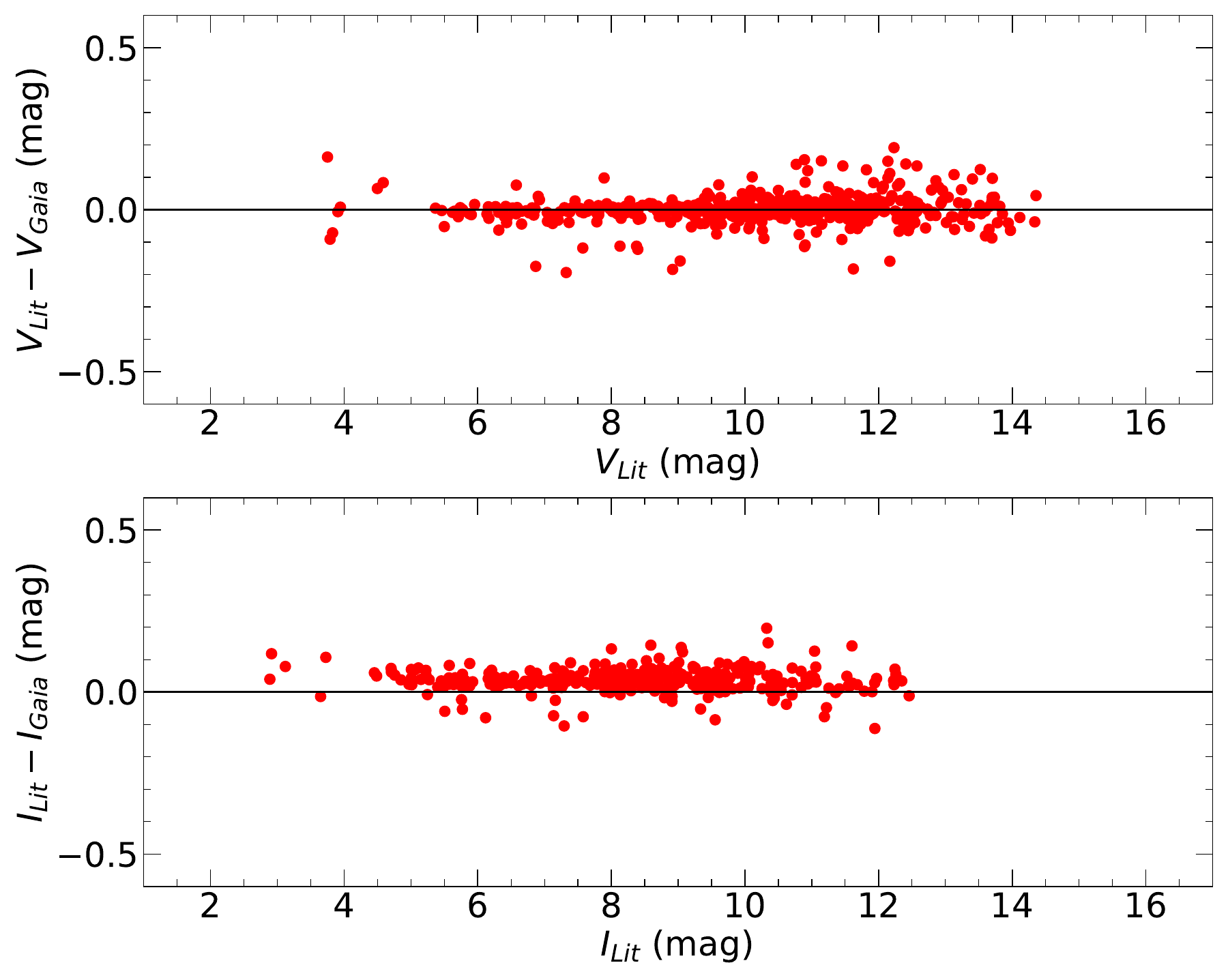}
   \caption{Comparison between the literature $V,\,I$ photometry and that from \gaia\ through the transformation by \citet{Pancino2022}. The top and bottom panels show the comparison in $V$ and $I$ bands, respectively. }
              \label{fig:photComparisonWithPancino}
   \end{figure}
   \begin{figure}
   \includegraphics[width=9cm]{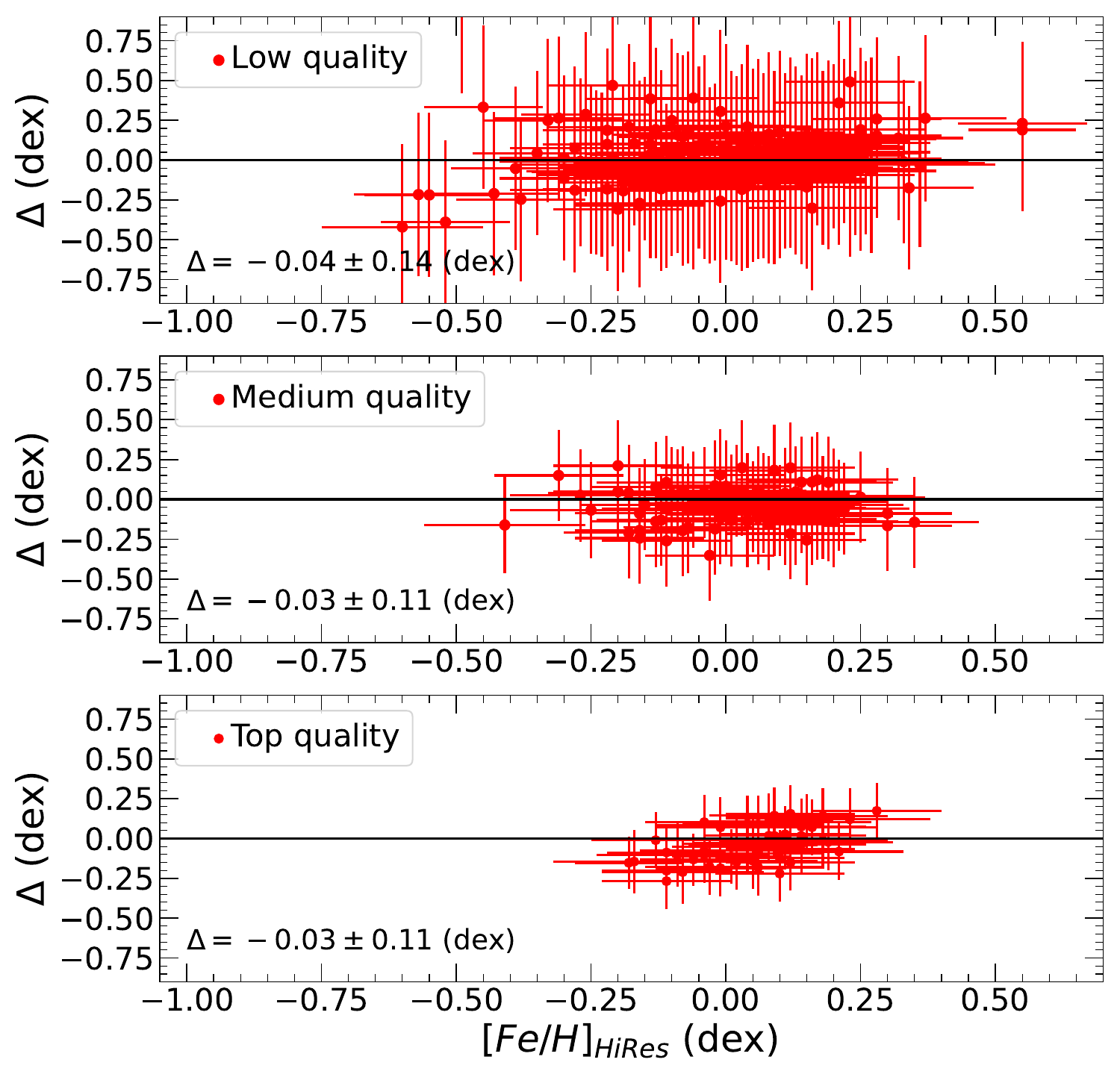}
   \caption{Comparison between the [Fe/H] from HiRes data and \gaia\ [M/H]. From top to bottom, the different panels show the comparison for the low-, medium-, and high-quality \gaia\ data, respectively (see text for details). In each panel, the average difference $\Delta$ and its dispersion is displayed.}
              \label{fig:metallicityComparison}
   \end{figure}
   
   \begin{figure}
   \includegraphics[width=9cm]{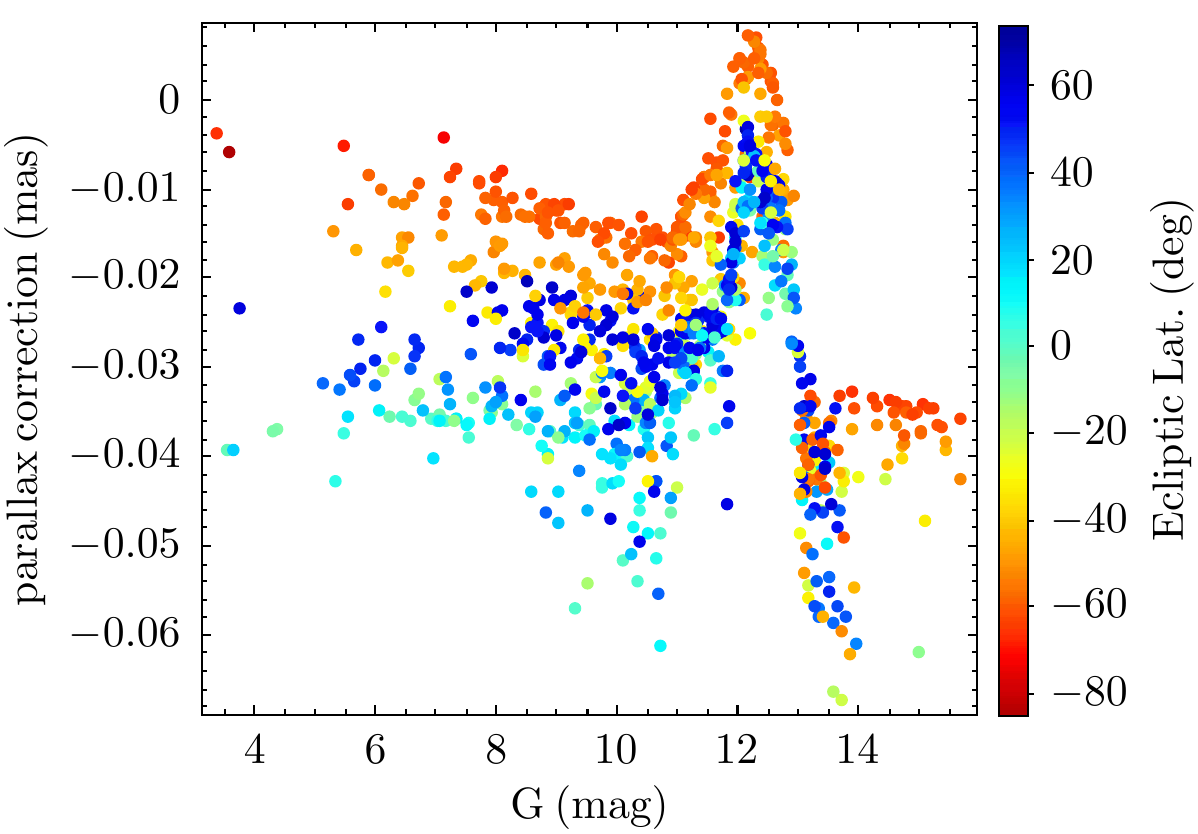}
   \caption{Individual parallax corrections from L21 as a function of the $G$ magnitude. The points are colour coded according to the ecliptic latitude.}
              \label{fig:parallaxCorrections}
   \end{figure}

\begin{sidewaystable*}
\caption{Photometric, astrometric, and spectroscopic data for the DCEP sample used in this work. Only the first ten lines of the table are shown here to guide the reader to its content. The machine-readable version of the table will be published at the CDS (Centre de Données astronomiques de Strasbourg, https://cds.u-strasbg.fr/). }\label{dataTable}
\footnotesize\setlength{\tabcolsep}{3pt}
\begin{tabular}{rlrrlrrrrrrrrrrrr}
\hline
  \multicolumn{1}{c}{GaiaDR3} &
  \multicolumn{1}{c}{Name} &
  \multicolumn{1}{c}{RA} &
  \multicolumn{1}{c}{Dec} &
  \multicolumn{1}{c}{Mode} &
  \multicolumn{1}{c}{plx} &
  \multicolumn{1}{c}{e\_plx} &
  \multicolumn{1}{c}{plx\_corr} &
  \multicolumn{1}{c}{RUWE} &
  \multicolumn{1}{c}{f\_v2} &
  \multicolumn{1}{c}{Period} &
  \multicolumn{1}{c}{$E(B-V)$} &
  \multicolumn{1}{c}{e\_$E(B-V)$} &
  \multicolumn{1}{c}{$G$} &
  \multicolumn{1}{c}{e\_$G$} &
  \multicolumn{1}{c}{\gbp} &
  \multicolumn{1}{c}{e\_\gbp} \\

    \multicolumn{1}{c}{} &
  \multicolumn{1}{c}{} &
  \multicolumn{1}{c}{(deg)} &
  \multicolumn{1}{c}{(deg)} &
  \multicolumn{1}{c}{} &
  \multicolumn{1}{c}{(mas)} &
  \multicolumn{1}{c}{(mas)} &
  \multicolumn{1}{c}{(mas)} &
  \multicolumn{1}{c}{} &
  \multicolumn{1}{c}{} &
  \multicolumn{1}{c}{(days)} &
  \multicolumn{1}{c}{(mag)} &
  \multicolumn{1}{c}{(mag)} &
  \multicolumn{1}{c}{(mag)} &
  \multicolumn{1}{c}{(mag)} &
  \multicolumn{1}{c}{(mag)} &
  \multicolumn{1}{c}{(mag)} \\

  \multicolumn{1}{c}{(1)} &
  \multicolumn{1}{c}{(2)} &
  \multicolumn{1}{c}{(3)} &
  \multicolumn{1}{c}{(4)} &
  \multicolumn{1}{c}{(5)} &
  \multicolumn{1}{c}{(6)} &
  \multicolumn{1}{c}{(7)} & 
  \multicolumn{1}{c}{(8)} &
  \multicolumn{1}{c}{(9)} &
  \multicolumn{1}{c}{(10)} &
  \multicolumn{1}{c}{(11)} &
  \multicolumn{1}{c}{(12)} &
  \multicolumn{1}{c}{(13)} &
  \multicolumn{1}{c}{(14)} &
  \multicolumn{1}{c}{(15)} &    
  \multicolumn{1}{c}{(16)} &
  \multicolumn{1}{c}{(17)} \\
  
\hline
\\
  3430067092837622272 & AA Gem & 91.64561  & 26.32922  & DCEP\_F & 0.2749 & 0.0201 & -0.0365 & 1.249 & 1.0 & 11.30157 & 0.345 & 0.036 &  9.363 & 0.005 &  9.956 & 0.007 \\
  3102535635624415872 & AA Mon & 104.34904 & -3.84334  & DCEP\_F & 0.3139 & 0.0171 & -0.0024 & 1.163 & 1.0 &  3.93815 & 0.765 & 0.017 & 12.103 & 0.004 & 12.993 & 0.009 \\
  4260210878780635904 & AA Ser & 280.34068 & -1.11122  & DCEP\_F & 0.2408 & 0.0335 & -0.0379 & 0.952 & 0.4 & 17.14211 & 1.295 & 0.139 & 10.910 & 0.002 & 12.485 & 0.004 \\
  473239154746762112  & AB Cam & 56.53439  & 58.78422  & DCEP\_F & 0.2128 & 0.0234 & -0.0278 & 1.276 & 1.0 &  5.78758 & 0.622 & 0.035 & 11.383 & 0.003 & 12.118 & 0.009 \\
  462252662762965120  & AC Cam & 50.94951  & 59.35567  & DCEP\_F & 0.3206 & 0.0202 & -0.0225 & 1.206 & 1.0 &  4.15677 & 0.904 & 0.090 & 11.830 & 0.003 & 12.805 & 0.049 \\
  3050050207554658048 & AC Mon & 105.24925 & -8.70899  & DCEP\_F & 0.3549 & 0.0211 & -0.0280 & 1.379 & 1.0 &  8.01493 & 0.507 & 0.033 &  9.599 & 0.006 & 10.314 & 0.005 \\
  462407693902385792  & AD Cam & 52.35821  & 60.44646  & DCEP\_F & 0.2965 & 0.0198 & -0.0185 & 1.185 & 1.0 & 11.26305 & 0.873 & 0.012 & 11.711 & 0.014 & 12.773 & 0.016 \\
  6057514092119497472 & AD Cru & 183.24861 & -62.09683 & DCEP\_F & 0.2929 & 0.0154 & -0.0224 & 1.017 & 1.0 &  6.39723 & 0.640 & 0.012 & 10.480 & 0.002 & 11.327 & 0.003 \\
  3378049163365268608 & AD Gem & 100.7813  & 20.93911  & DCEP\_F & 0.3356 & 0.0223 & -0.0342 & 0.969 & 1.0 &  3.78800 & 0.206 & 0.048 &  9.648 & 0.009 & 10.021 & 0.018 \\
  5614312705966204288 & AD Pup & 117.01605 & -25.57778 & DCEP\_F & 0.2331 & 0.0189 & -0.0206 & 1.362 & 1.0 & 13.59681 & 0.363 & 0.020 &  9.541 & 0.003 & 10.125 & 0.015 \\
\hline
\\
\end{tabular}
\begin{tabular}{rrrrrrrrrrrrrcccccc}
\hline
  \multicolumn{1}{c}{\grp} &
  \multicolumn{1}{c}{e\_\grp} &
  \multicolumn{1}{c}{$V$} &
  \multicolumn{1}{c}{e\_$V$} &
  \multicolumn{1}{c}{$I$} &
  \multicolumn{1}{c}{e\_$I$} &
  \multicolumn{1}{c}{$J$} &
  \multicolumn{1}{c}{e\_$J$} &
  \multicolumn{1}{c}{$H$} &
  \multicolumn{1}{c}{e\_$H$} &
  \multicolumn{1}{c}{$K_s$} &
  \multicolumn{1}{c}{e\_$K_s$} &
  \multicolumn{1}{c}{[Fe/H]} &
  \multicolumn{1}{c}{e\_[Fe/H]} &
  \multicolumn{1}{c}{metSource} &
  \multicolumn{1}{c}{ebvSource} &
  \multicolumn{1}{c}{vSource} &
  \multicolumn{1}{c}{iSource} &
  \multicolumn{1}{c}{sflag} \\

    \multicolumn{1}{c}{(mag)} &
  \multicolumn{1}{c}{(mag)} &
  \multicolumn{1}{c}{(mag)} &
  \multicolumn{1}{c}{(mag)} &
  \multicolumn{1}{c}{(mag)} &
  \multicolumn{1}{c}{(mag)} &
  \multicolumn{1}{c}{(mag)} &
  \multicolumn{1}{c}{(mag)} &
  \multicolumn{1}{c}{(mag)} &
  \multicolumn{1}{c}{(mag)} &
  \multicolumn{1}{c}{(mag)} &
  \multicolumn{1}{c}{(mag)} &
  \multicolumn{1}{c}{(dex)} &
  \multicolumn{1}{c}{(dex)} &
  \multicolumn{1}{c}{} &
  \multicolumn{1}{c}{} &
  \multicolumn{1}{c}{} &
  \multicolumn{1}{c}{} &
  \multicolumn{1}{c}{} \\

  \multicolumn{1}{c}{(18)} &
  \multicolumn{1}{c}{(19)} &
  \multicolumn{1}{c}{(20)} &
  \multicolumn{1}{c}{(21)} &
  \multicolumn{1}{c}{(22)} &
  \multicolumn{1}{c}{(23)} &
  \multicolumn{1}{c}{(24)} & 
  \multicolumn{1}{c}{(25)} &
  \multicolumn{1}{c}{(26)} &
  \multicolumn{1}{c}{(27)} &
  \multicolumn{1}{c}{(28)} &
  \multicolumn{1}{c}{(29)} &
  \multicolumn{1}{c}{(30)} &
  \multicolumn{1}{c}{(31)} &
  \multicolumn{1}{c}{(32)} &    
  \multicolumn{1}{c}{(33)} &
  \multicolumn{1}{c}{(34)} &    
  \multicolumn{1}{c}{(35)} &
  \multicolumn{1}{c}{(36)} \\
    
  \hline
  \\
   8.620 & 0.005 &  9.722 & 0.020 &  8.581 & 0.020 & 7.636 & 0.008 & 7.201 & 0.008 & 7.048 & 0.008 & -0.08 & 0.12 & G18 & G18 & Lit & Lit & 0\\        
  11.174 & 0.004 & 12.744 & 0.020 & 11.085 & 0.020 & 9.729 & 0.008 & 9.199 & 0.008 & 8.953 & 0.008 & -0.12 & 0.12 & G18 & G18 & Lit & Lit & 0\\        
   9.716 & 0.005 & 12.243 & 0.020 &  9.571 & 0.020 & 7.563 & 0.008 & 6.773 & 0.008 & 6.457 & 0.008 &  0.38 & 0.12 & G18 & G18 & Lit & Lit & 0\\        
  10.533 & 0.007 & 11.860 & 0.020 & 10.454 & 0.030 & 9.418 & 0.025 & 8.936 & 0.025 & 8.746 & 0.025 & -0.11 & 0.12 & G18 & G18 & Lit & GDR3+P22 & 0\\   
  10.822 & 0.009 & 12.605 & 0.020 & 10.697 & 0.020 & 9.325 & 0.025 & 8.752 & 0.025 & 8.499 & 0.025 & -0.16 & 0.12 & G18 & G18 & Lit & Lit & 0\\        
   8.771 & 0.004 & 10.099 & 0.020 &  8.707 & 0.020 & 7.577 & 0.008 & 7.066 & 0.008 & 6.849 & 0.008 & -0.06 & 0.12 & G18 & G18 & Lit & Lit & 0\\        
  10.690 & 0.010 & 12.560 & 0.020 & 10.587 & 0.030 & 8.988 & 0.025 & 8.363 & 0.025 & 8.109 & 0.025 & -0.28 & 0.12 & G18 & G18 & Lit & GDR3+P22 & 0\\   
   9.569 & 0.002 & 11.061 & 0.020 &  9.452 & 0.020 & 8.198 & 0.025 & 7.661 & 0.025 & 7.387 & 0.025 &  0.08 & 0.12 & G18 & G18 & Lit & Lit & 0\\        
   9.108 & 0.010 &  9.855 & 0.020 &  9.050 & 0.020 & 8.448 & 0.008 & 8.151 & 0.008 & 8.027 & 0.008 & -0.14 & 0.12 & G18 & G18 & Lit & Lit & 0\\        
   8.786 & 0.008 &  9.919 & 0.020 &  8.736 & 0.020 & 7.723 & 0.025 & 7.341 & 0.025 & 7.144 & 0.025 & -0.06 & 0.12 & G18 & G18 & Lit & Lit & 0\\
\hline \hline
\end{tabular}
\tablefoot{The meaning of the different columns is as follows: (1) \gaia\ EDR3 identification; (2) other names of the DCEPs; (3) and (4) equatorial coordinates (J2000); (5) mode of pulsation -- F, 1O, F1O, and 1O2O indicate the fundamental, first overtone, and the mixed-mode pulsation modes, respectively; (6) and (7) original parallax value and error from \gaia\ EDR3 catalogue; (8) parallax 
 correction according to \citet{Lindegren2021}; (9) RUWE value from \gaia\ EDR3; (10) Fidelity\_v2 index from \citet{Rybizki2022}; (11) period of pulsation. For mixed-mode DCEPs, the longest period is listed; (12) and (13) E(B-V) and error; (14) to (29) magnitudes and errors for the $G$\,\gbp\,\grp\,\,$V$,\,$I$,\,$J$,\,$H$,\,$K_s$ bands, respectively; (30) and (31) iron abundance and error; (32) literature source of the iron abundance -- G14=\citet{Genovali2014}, GALAH = \citet{Galah2021}; GDR3 = \gaia\ DR3 \citep[see][]{RecioBlanco2023}; GC17=\citet{Gaia2017}, G18=\citet{Groenewegen2018}, K22=\citet{Kovtyukh2022}; PASTEL= \citet{Pastel2016};   R21=\citet{Ripepi2021a}; T23=\citet{Trentin2023}; (33) source for the values of $E(B-V)$ (in addition to those quoted for the metallicity) -- VL07 = \citet{vanLeeuwen2007}; R21 means that the value of reddening was calculated using the period--colour relations published in \citet{Ripepi2021a}; (34) and (35) source for the $V$ and $I$ magnitudes -- Lit = compilation from the literature (mainly from G18); Gaia+P22 = magnitudes calculated on the basis of the \gaia\ bands using the transformations to the Johnson-Cousins system by \citet{Pancino2022}; (36) flag related to the provenence of the [Fe/H] value: 0 = HiRes spectroscopy; 1 and 2 = \gaia\ RVS spectroscopy published in DR3 with good- and intermediate-quality according to the selections by \citet{RecioBlanco2023}.}
\end{sidewaystable*}

\begin{table}
    \centering
    \caption{Photometric bands and colour coefficients adopted to calculate the Wesenheit magnitudes considered in this work.}
    \begin{tabular}{l|c}
    \hline
    \hline
    Bands&$\xi$\\
    \hline
    $\rm W_{G,G_{BP}-G_{RP}}$ & 1.9\\
    $\rm W_{H,V-I}$ & 0.461\\
    $\rm W_{H,V-I}^{cHST}$ & 0.386\\
    $\rm W_{K_S,V-K_S}$ & 0.130\\
    $\rm W_{K_S,J-K_S}$ & 0.690\\
    \hline
    \end{tabular}
    \tablefoot{The Wesenhit magnitude names are contained in column 1, while the $\xi$ values from the Cardelli law \citep[][]{Cardelli1989} are in column 2.}
    \label{tab-wesCoeff}
\end{table}

\begin{table*}
\footnotesize\setlength{\tabcolsep}{3pt}
\caption{Results of the fitting using the Lit.+ Gaia data set.} 
\label{tab:fitResSflag2}
\centering
\begin{tabular}{cccccccccccccc}
  \hline
ID & $\alpha$ & $\beta$ & $\gamma$ & $\delta$ & $\mu_0^{LMC}$ & $rms$ &  Band & Mode & $N_{dat}$   \\ 
  \hline
 1 & $-4.248 \pm 0.025$ & $-2.566 \pm 0.063$ & $-0.611 \pm 0.121$ & $        -       $ & $18.398 \pm 0.058$ & 0.022 & $PLG$                  & F    & 481 \\ 
 2 & $-4.255 \pm 0.028$ & $-2.620 \pm 0.066$ & $-0.571 \pm 0.133$ & $ 0.615 \pm 0.363$ & $18.294 \pm 0.071$ & 0.024 & $PLG$                  & F    & 485 \\ 
 3 & $-4.295 \pm 0.022$ & $-2.529 \pm 0.053$ & $-0.604 \pm 0.083$ & $        -       $ & $18.435 \pm 0.034$ & 0.027 & $PLG$                  & F+1O & 754 \\ 
 4 & $-4.298 \pm 0.023$ & $-2.527 \pm 0.052$ & $-0.576 \pm 0.114$ & $ 0.120 \pm 0.213$ & $18.426 \pm 0.035$ & 0.027 & $PLG$                  & F+1O & 753 \\ 
 5 & $-3.867 \pm 0.028$ & $-2.344 \pm 0.064$ & $-0.487 \pm 0.117$ & $        -       $ & $18.422 \pm 0.056$ & 0.025 & $PLG_{BP}$             & F    & 467 \\ 
 6 & $-3.882 \pm 0.031$ & $-2.375 \pm 0.068$ & $-0.385 \pm 0.146$ & $ 0.213 \pm 0.344$ & $18.427 \pm 0.067$ & 0.026 & $PLG_{BP}$             & F    & 470 \\ 
 7 & $-3.905 \pm 0.024$ & $-2.250 \pm 0.055$ & $-0.488 \pm 0.086$ & $        -       $ & $18.459 \pm 0.038$ & 0.030 & $PLG_{BP}$             & F+1O & 743 \\ 
 8 & $-3.907 \pm 0.026$ & $-2.252 \pm 0.056$ & $-0.482 \pm 0.113$ & $ 0.057 \pm 0.190$ & $18.450 \pm 0.032$ & 0.030 & $PLG_{BP}$             & F+1O & 746 \\ 
 9 & $-4.696 \pm 0.017$ & $-2.665 \pm 0.051$ & $-0.563 \pm 0.085$ & $        -       $ & $18.346 \pm 0.040$ & 0.014 & $PLG_{RP}$             & F    & 458 \\ 
10 & $-4.700 \pm 0.018$ & $-2.690 \pm 0.057$ & $-0.556 \pm 0.100$ & $ 0.315 \pm 0.327$ & $18.300 \pm 0.058$ & 0.014 & $PLG_{RP}$             & F    & 457 \\ 
11 & $-4.752 \pm 0.020$ & $-2.614 \pm 0.046$ & $-0.506 \pm 0.069$ & $        -       $ & $18.423 \pm 0.031$ & 0.018 & $PLG_{RP}$             & F+1O & 739 \\ 
12 & $-4.772 \pm 0.022$ & $-2.684 \pm 0.052$ & $-0.386 \pm 0.100$ & $ 0.353 \pm 0.193$ & $18.391 \pm 0.030$ & 0.018 & $PLG_{RP}$             & F+1O & 744 \\ 
13 & $-5.663 \pm 0.016$ & $-3.087 \pm 0.048$ & $-0.511 \pm 0.078$ & $        -       $ & $18.382 \pm 0.034$ & 0.010 & $PLH$                  & F    & 488 \\ 
14 & $-5.668 \pm 0.016$ & $-3.106 \pm 0.050$ & $-0.478 \pm 0.087$ & $ 0.208 \pm 0.275$ & $18.354 \pm 0.046$ & 0.010 & $PLH$                  & F    & 487 \\ 
15 & $-5.699 \pm 0.013$ & $-3.070 \pm 0.036$ & $-0.391 \pm 0.056$ & $        -       $ & $18.465 \pm 0.022$ & 0.011 & $PLH$                  & F+1O & 742 \\ 
16 & $-5.699 \pm 0.013$ & $-3.067 \pm 0.035$ & $-0.396 \pm 0.065$ & $-0.014 \pm 0.110$ & $18.466 \pm 0.023$ & 0.011 & $PLH$                  & F+1O & 742 \\ 
17 & $-4.774 \pm 0.017$ & $-2.676 \pm 0.052$ & $-0.602 \pm 0.084$ & $        -       $ & $18.361 \pm 0.039$ & 0.013 & $PLI$                  & F    & 460 \\ 
18 & $-4.779 \pm 0.018$ & $-2.722 \pm 0.058$ & $-0.561 \pm 0.096$ & $ 0.482 \pm 0.308$ & $18.278 \pm 0.055$ & 0.014 & $PLI$                  & F    & 464 \\ 
19 & $-4.830 \pm 0.019$ & $-2.632 \pm 0.045$ & $-0.507 \pm 0.065$ & $        -       $ & $18.452 \pm 0.029$ & 0.018 & $PLI$                  & F+1O & 735 \\ 
20 & $-4.837 \pm 0.020$ & $-2.643 \pm 0.044$ & $-0.445 \pm 0.091$ & $ 0.226 \pm 0.168$ & $18.433 \pm 0.026$ & 0.017 & $PLI$                  & F+1O & 736 \\ 
21 & $-5.343 \pm 0.017$ & $-2.945 \pm 0.052$ & $-0.575 \pm 0.086$ & $        -       $ & $18.361 \pm 0.040$ & 0.011 & $PLJ$                  & F    & 485 \\ 
22 & $-5.344 \pm 0.018$ & $-2.948 \pm 0.055$ & $-0.564 \pm 0.096$ & $ 0.036 \pm 0.296$ & $18.357 \pm 0.054$ & 0.011 & $PLJ$                  & F    & 482 \\ 
23 & $-5.380 \pm 0.014$ & $-2.888 \pm 0.038$ & $-0.503 \pm 0.061$ & $        -       $ & $18.438 \pm 0.025$ & 0.014 & $PLJ$                  & F+1O & 755 \\ 
24 & $-5.380 \pm 0.014$ & $-2.888 \pm 0.038$ & $-0.498 \pm 0.074$ & $ 0.012 \pm 0.121$ & $18.436 \pm 0.025$ & 0.014 & $PLJ$                  & F+1O & 756 \\ 
25 & $-5.766 \pm 0.016$ & $-3.145 \pm 0.046$ & $-0.527 \pm 0.080$ & $        -       $ & $18.374 \pm 0.037$ & 0.009 & $PLK$                  & F    & 477 \\ 
26 & $-5.773 \pm 0.018$ & $-3.179 \pm 0.052$ & $-0.481 \pm 0.088$ & $ 0.350 \pm 0.266$ & $18.324 \pm 0.050$ & 0.009 & $PLK$                  & F    & 482 \\ 
27 & $-5.793 \pm 0.014$ & $-3.121 \pm 0.032$ & $-0.458 \pm 0.052$ & $        -       $ & $18.420 \pm 0.024$ & 0.010 & $PLK$                  & F+1O & 743 \\ 
28 & $-5.796 \pm 0.014$ & $-3.125 \pm 0.031$ & $-0.410 \pm 0.066$ & $ 0.203 \pm 0.117$ & $18.400 \pm 0.023$ & 0.010 & $PLK$                  & F+1O & 741 \\ 
29 & $-3.975 \pm 0.018$ & $-2.354 \pm 0.057$ & $-0.423 \pm 0.090$ & $        -       $ & $18.415 \pm 0.045$ & 0.021 & $PLV$                  & F    & 460 \\ 
30 & $-3.976 \pm 0.019$ & $-2.362 \pm 0.058$ & $-0.404 \pm 0.108$ & $ 0.106 \pm 0.330$ & $18.406 \pm 0.053$ & 0.021 & $PLV$                  & F    & 459 \\ 
31 & $-4.033 \pm 0.022$ & $-2.317 \pm 0.055$ & $-0.364 \pm 0.076$ & $        -       $ & $18.490 \pm 0.036$ & 0.026 & $PLV$                  & F+1O & 738 \\ 
32 & $-4.050 \pm 0.024$ & $-2.380 \pm 0.054$ & $-0.261 \pm 0.115$ & $ 0.458 \pm 0.226$ & $18.420 \pm 0.034$ & 0.028 & $PLV$                  & F+1O & 746 \\ 
33 & $-5.917 \pm 0.017$ & $-3.245 \pm 0.055$ & $-0.745 \pm 0.085$ & $        -       $ & $18.310 \pm 0.036$ & 0.009 & $PW_{G,G_{BP}-G_{RP}}$ & F    & 478 \\ 
34 & $-5.927 \pm 0.018$ & $-3.292 \pm 0.044$ & $-0.659 \pm 0.100$ & $ 0.559 \pm 0.273$ & $18.254 \pm 0.048$ & 0.009 & $PW_{G,G_{BP}-G_{RP}}$ & F    & 484 \\ 
35 & $-5.960 \pm 0.018$ & $-3.230 \pm 0.041$ & $-0.573 \pm 0.066$ & $        -       $ & $18.436 \pm 0.028$ & 0.010 & $PW_{G,G_{BP}-G_{RP}}$ & F+1O & 726 \\ 
36 & $-5.958 \pm 0.017$ & $-3.221 \pm 0.038$ & $-0.583 \pm 0.089$ & $-0.036 \pm 0.174$ & $18.439 \pm 0.028$ & 0.010 & $PW_{G,G_{BP}-G_{RP}}$ & F+1O & 725 \\ 
37 & $-6.042 \pm 0.016$ & $-3.239 \pm 0.051$ & $-0.551 \pm 0.082$ & $        -       $ & $18.385 \pm 0.034$ & 0.008 & $PW_{H,V-I}$           & F    & 487 \\ 
38 & $-6.050 \pm 0.015$ & $-3.279 \pm 0.039$ & $-0.502 \pm 0.079$ & $ 0.390 \pm 0.250$ & $18.333 \pm 0.047$ & 0.008 & $PW_{H,V-I}$           & F    & 491 \\ 
39 & $-6.073 \pm 0.013$ & $-3.239 \pm 0.035$ & $-0.447 \pm 0.058$ & $        -       $ & $18.457 \pm 0.023$ & 0.010 & $PW_{H,V-I}$           & F+1O & 747 \\ 
40 & $-6.072 \pm 0.014$ & $-3.231 \pm 0.035$ & $-0.444 \pm 0.067$ & $-0.026 \pm 0.115$ & $18.464 \pm 0.023$ & 0.010 & $PW_{H,V-I}$           & F+1O & 746 \\ 
41 & $-5.908 \pm 0.014$ & $-3.167 \pm 0.040$ & $-0.338 \pm 0.067$ & $        -       $ & $18.421 \pm 0.032$ & 0.006 &  $PW_{H,V-I}^{cHST}$   & F    & 369 \\ 
42 & $-5.903 \pm 0.015$ & $-3.207 \pm 0.049$ & $-0.366 \pm 0.078$ & $ 0.366 \pm 0.259$ & $18.325 \pm 0.047$ & 0.007 &  $PW_{H,V-I}^{cHST}$   & F    & 393 \\ 
43 & $-5.915 \pm 0.013$ & $-3.133 \pm 0.036$ & $-0.369 \pm 0.063$ & $        -       $ & $18.414 \pm 0.028$ & 0.008 &  $PW_{H,V-I}^{cHST}$   & F+1O & 491 \\ 
44 & $-5.922 \pm 0.013$ & $-3.156 \pm 0.035$ & $-0.322 \pm 0.071$ & $ 0.267 \pm 0.110$ & $18.369 \pm 0.026$ & 0.008 &  $PW_{H,V-I}^{cHST}$   & F+1O & 499 \\ 
45 & $-6.064 \pm 0.016$ & $-3.286 \pm 0.045$ & $-0.494 \pm 0.074$ & $        -       $ & $18.378 \pm 0.035$ & 0.008 & $PW_{K_S,J-K_S}$       & F    & 480 \\ 
46 & $-6.070 \pm 0.015$ & $-3.336 \pm 0.041$ & $-0.470 \pm 0.080$ & $ 0.482 \pm 0.231$ & $18.283 \pm 0.045$ & 0.008 & $PW_{K_S,J-K_S}$       & F    & 481 \\ 
47 & $-6.085 \pm 0.015$ & $-3.294 \pm 0.035$ & $-0.446 \pm 0.049$ & $        -       $ & $18.407 \pm 0.022$ & 0.009 & $PW_{K_S,J-K_S}$       & F+1O & 737 \\ 
48 & $-6.089 \pm 0.015$ & $-3.292 \pm 0.035$ & $-0.389 \pm 0.070$ & $ 0.197 \pm 0.138$ & $18.395 \pm 0.023$ & 0.009 & $PW_{K_S,J-K_S}$       & F+1O & 731 \\ 
49 & $-6.016 \pm 0.016$ & $-3.261 \pm 0.047$ & $-0.507 \pm 0.077$ & $        -       $ & $18.386 \pm 0.035$ & 0.008 & $PW_{K_S,V-K_S}$       & F    & 477 \\ 
50 & $-6.020 \pm 0.017$ & $-3.285 \pm 0.052$ & $-0.471 \pm 0.086$ & $ 0.269 \pm 0.265$ & $18.350 \pm 0.048$ & 0.008 & $PW_{K_S,V-K_S}$       & F    & 481 \\ 
51 & $-6.030 \pm 0.013$ & $-3.239 \pm 0.032$ & $-0.493 \pm 0.052$ & $        -       $ & $18.405 \pm 0.022$ & 0.009 & $PW_{K_S,V-K_S}$       & F+1O & 735 \\ 
52 & $-6.034 \pm 0.013$ & $-3.242 \pm 0.031$ & $-0.440 \pm 0.065$ & $ 0.164 \pm 0.114$ & $18.396 \pm 0.023$ & 0.009 & $PW_{K_S,V-K_S}$       & F+1O & 732 \\ 
   \hline
   \hline
\end{tabular}
\tablefoot{The ID identifies each different fit to the data; $\alpha,\beta,\gamma \,and \, \delta$ are the coefficients of the $PLZ/PWZ$ relations; $\mu_0^{LMC}$ represents the {true} distance modulus of the LMC; $rms$ is the root mean square; Band specifies the photometric band considered in the fit; Mode identifies the sample adopted; $N_{dat}$ is the number of DCEPs adopted.}
\end{table*}

Direct empirical evaluations of the metallicity dependence of $PL$ relations using Galactic DCEPs 
with sound [Fe/H] measurements based on high-resolution (HiRes hereafter) spectroscopy have been hampered so far by the lack of accurate independent distances for a significant number of Milky Way (MW) DCEPs. The advent of the \gaia\ mission \citep[][]{Gaia2016} has completely changed this scenario.   \gaia\ began providing accurate parallaxes with data release 2 \citep[DR2][]{Gaia2018}, which were further improved with the early data release 3 \citep[][]{Gaia2021}. In addition, the \gaia\ mission secured the discovery of hundreds of new Galactic DCEPs \citep[][]{Clementini2019,Ripepi2019,Ripepi2022a} which, together with those discovered by other surveys, such as the OGLE Galactic Disk survey \citep[][]{Udalski2018} and the Zwicky Transient Facility \citep[ZTF][]{Chen2020}, constitute a formidable sample of DCEPs useful not only in the context of the extragalactic distance scale but also for Galactic studies \citep[e.g.][and references therein]{Lemasle2022,Trentin2023}.   
However, until a few years ago, the number of DCEPs with metallicity measurements from HiRes spectroscopy was mainly restricted to the solar neighbourhood, where the DCEPs span a limited range in [Fe/H], centred on solar or slightly supersolar values, with a small dispersion of 0.2-0.3 dex \cite[e.g.][]{Genovali2014,Luck2018,Groenewegen2018,Ripepi2019}. This makes it very difficult to measure the metallicity dependence of $PL$ relations using Galactic Cepheids  with sound statistical significance.  

In this context, a few years ago we started a project named C-MetaLL \citep[Cepheid - Metallicity in the Leavitt Law, see][for a full description]{Ripepi2021a}, with the goal being to measure the chemical abundance of a sample of 250-300 Galactic DCEPs through HiRes spectroscopy, expressly aiming to enlarge the iron abundance range towards the metal-poor regime ---that is, [Fe/H]$<-$0.4 dex--- where only a few stars have abundance measurements in the literature. In the first two papers of the series \citep[][]{Ripepi2021a,Trentin2023}, we published accurate abundances for more than 25 chemical species for a total of 114 DCEPs. In particular, in \citet{Trentin2023}, we obtained measures for 43 objects with [Fe/H]$<-$0.4 dex, reaching abundances of as low as $-$1.1 dex. 
The scope of this paper is to study the metallicity dependence of the $PL$ relations using a [Fe/H] range of more than 1 dex. 
In this way, we aim to be able to discern between the two scenarios that came out in the recent literature concerning the metallicity dependence of $PL$ relations. In fact, in our previous works using Galactic DCEPs and \gaia\ parallaxes, we found a rather large dependence in the NIR bands of on the order of $\sim -0.4$ mag/dex \citep[][]{Ripepi2020,Ripepi2021a}; or even larger when using the \gaia\ bands \citep[$\sim -0.5$ mag/dex][]{Ripepi2022a}. These values are discrepant with those measured by the SH0ES group \citep[$\sim -0.2$ mag/dex][]{Riess2021} ---which, on the other hand, used a calibrating sample of 75 DCEPs in the solar vicinity spanning a small [Fe/H] range---, and are also different from those measured by \citet{Breuval2021,Breuval2022}, who obtained similar results to the SH0ES group using three Cepheid samples in the Milky Way and the Large and Small Magellanic Clouds (LMC and SMC, respectively) as three representative objects with different mean metallicities.

The paper is organized as follows: in Sect. \ref{sec:data} we describe the sample of DCEPs and their properties. In Sect.~\ref{sec:method}, we describe the method we used to derive the period--luminosity--metallicity ($PLZ$) and period--Wesenheit--metallicity ($PWZ$) relations; in Sect. \ref{sec:results} and Sect. \ref{sec:discussion}, we describe and discuss our results; and in Sect. \ref{sec:conclusion} we outline our conclusions.


\section{Description of the data used in this work}\label{sec:data}

In this section, we describe the sample of DCEPs used in this paper and their photometric, spectroscopic, and astrometric properties.  
All the data employed in our analysis are listed in Table~\ref{dataTable}.

\subsection{Photometry}
\label{Sect:photometry}

Optical $V,\,I$\footnote{The $I$ photometry is in the Cousins system} photometry is available in the literature for about 488 and 364 DCEPs of our sample, respectively \citep[][]{Groenewegen2018,Ripepi2021a}. For the remaining stars, we decided to use the homogeneous and precise \gaia\ $G$, \gbp, \grp\ photometry transformed into the Johnson-Cousins $V,\,I$ bands by means of the relations published by \citet{Pancino2022}. We calculated these aforementioned magnitudes for all the stars with full $G$, \gbp, \grp\ values (all DCEPs except one) using the intensity-averaged magnitudes from the \gaia\ Vizier catalogue \textit{I/358/vcep} \citep[][]{Ripepi2022a} or the simple average magnitudes from the \gaia\ source catalogue (Vizier I/355/gaiadr3) for the few stars not present in the quoted catalogue. 
Figure~\ref{fig:photComparisonWithPancino} shows the comparison between the literature $V,\,I$ magnitudes and those calculated from the \gaia\ photometry for the stars for which both values are available. While the transformed $V$ photometry appears perfectly compatible with that from the literature ($V_{Lit}-V_{Gaia}\sim 0.0$ with dispersion $\sigma \sim 0.04$ mag), the transformed $I$ photometry appears to be slightly too faint: $I_{Lit}-I_{Gaia} \sim 0.035$ (mag) with a dispersion of $\sigma \sim 0.035$ mag. Therefore, we used the transformed $V$ bands with no modification, while we corrected the transformed $I$ bands by increasing their value by 0.035 mag. 
As for the uncertainties, we assumed 0.02 mag in both $V$ and $I$ for the literature sample when the number is not available in the original publication, while for the remaining stars, we propagated the errors, also taking into account (summing in quadrature) the uncertainties in the transformations provided by \citet{Pancino2022}. For a handful of stars, the DCEPs (\gbp-\grp) colours were beyond the validity limit of the  \citet{Pancino2022} relations. For these stars, we adopted the synthetic $V$, $I$ magnitudes calculated by \citet{Montegriffo2023} based on \gaia\ DR3 photometry.   

Near-infrared $J,\,H,$ and $\,K_S$ band photometry is from \citet{vanLeeuwen2007}, \citet{Gaia2017}, \citet{Groenewegen2018} and \citet{Ripepi2021a} for the literature sample, and is derived from single-epoch 2MASS photometry \citep{Skrutskie2006} for the remaining stars. To this aim, we adopted the procedure outlined in \citet{Ripepi2021a}, using the ephemerides (periods and epochs of maximum light) from the \gaia\ Vizier catalogue \textit{I/358/vcep}. As in our previous work, the uncertainties on the mean magnitudes were calculated using Monte Carlo simulations, varying the 2MASS magnitude and the phase of the single-epoch photometry within their errors, where the last quantity was calculated using the errors on the periods given in the \gaia\ catalogue.    

In this work, we also considered the Wesenheit indices, which are reddening-free quantities by construction \citep{Madore1976}. These are obtained by combining the standard magnitude in a given photometric band with a colour term according to the following equation:
\begin{equation}
    W_{X_1,X_2-X_3} = X_1 -\xi^{2;3}_1\cdot(X_2 - X_3),
\end{equation}
where $X_i$ indicates the generic band and the coefficient $\xi^{2,3}_1$ coincides with the total-to-selective absorption and is obtained by assuming an extinction law. The photometric band combinations adopted in this work are listed in Table~\ref{tab-wesCoeff},
together with the coefficient of the colour term. In order to transform the Johnson-Cousins-2MASS ground-based $\rm H,\,V, and\,I$ photometry into the HST correspondent F160W,\,F555W, and\,F814W filters, we considered the photometric transformations by \citet{Riess2021}.
 We note that, for brevity, in the following the calibrated Wesenheit in the HST bands is referred to as $W_{H,V-I}^{cHST}$.\footnote{ We caution the reader that the conversion equations in \citet{Riess2021} contain a typo regarding the F814W filter. In this work we used the correct relation $F814W = V - $0.48$ (V-H) - 0.025$ instead of $F814W = V - 0.48 (J-H) - 0.025$}.


In analogy with photometry, the reddening for the literature sample was taken from \citet{vanLeeuwen2007}, \citet{Gaia2017}, \citet{Groenewegen2018} and \citet{Ripepi2021a}, while for the remaining stars, we used the same period--colour relations used in \citet{Ripepi2021a} involving $(V-I)$ colour (their eq. 4), obtaining $E(V-I)$ excesses that were converted into the corresponding $E(B-V)$ ones using the relation $E(V-I)$=1.28 $E(B-V)$ \citep[][]{Tammann2003}. The uncertainties on these reddening values were calculated by summing in quadrature the rms of eq. 4 in \citet{Ripepi2021a}, with the errors on $V$ and $I$ magnitudes. However, to be conservative, we assumed an uncertainty of 10\% on the reddening  for
values of $E(B-V)>1.0$ mag.   


   \begin{figure}
   \includegraphics[width=9cm]{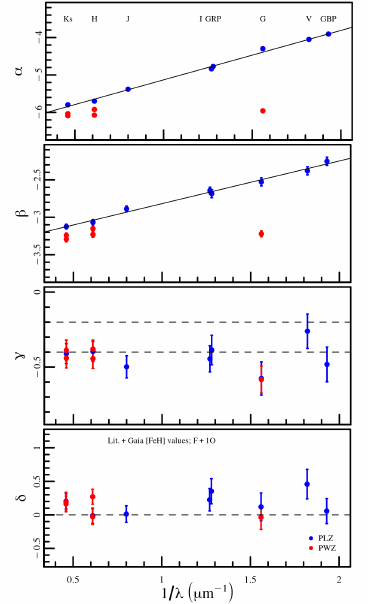}
   \caption{Coefficients of $PLZ/PWZ$ relations obtained from the fit to the Lit. + Gaia sample plotted against the $\lambda^{-1}$ parameter. From top to bottom, the panels correspond to the $\alpha$, $\beta$, $\gamma,$ and $\delta$ coefficients. In each panel, $PLZ$ and $PWZ$ coefficients are plotted with blue and red symbols respectively, while the solid line shows the linear fit only for the $PLZ$ relations. Grey dashed lines in the $\gamma$ panel delimit the range of results in the literature \citep[$-0.2$ to $-0.4$ mag/dex, see ][and Sect.~\ref{sec:comparison}]{Breuval2022}, while in the $\delta$ panel, it corresponds to the value 0.}
              \label{fig:coeffs-vs-wlen-abcd-F1O-sflag2}
   \end{figure}
   
   \begin{figure}
   \includegraphics[width=9cm]{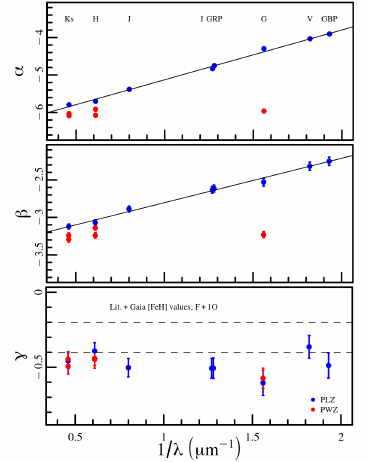}
   \caption{Same as Fig.~\ref{fig:coeffs-vs-wlen-abcd-F1O-sflag2} but for the case neglecting the metallicity dependence in the PLZ/PLW slope (i.e. $\delta=0$).}
              \label{fig:coeffs-vs-wlen-abc-F1O-sflag2}
   \end{figure}

   \begin{figure*}
   \includegraphics[width=18cm]{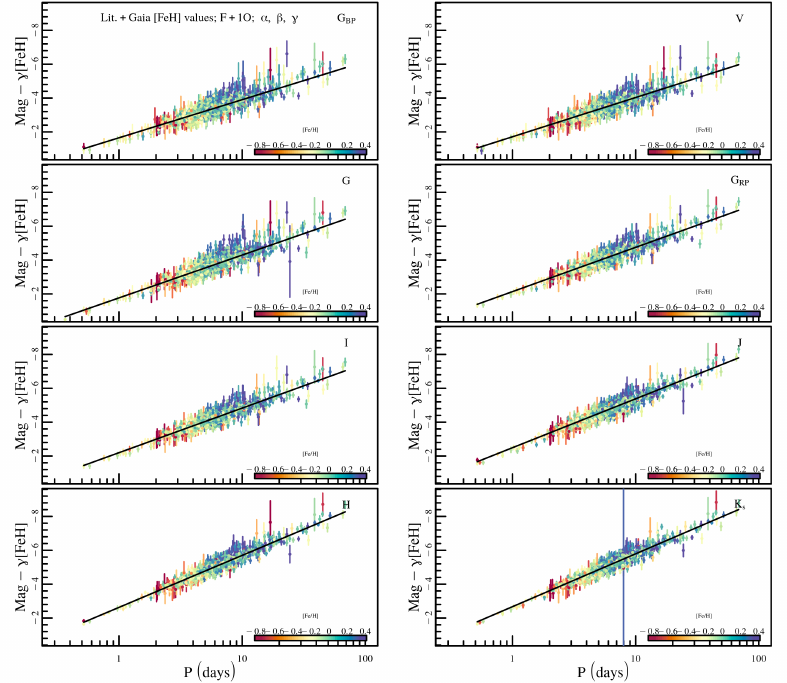}
   \caption{PL relations for the bands studied in the paper where the magnitudes have been subtracted by the metallicity contribution. The colour bar represents the metallicity.}
              \label{fig:AbsMagFeH_logP_abc_F1O_sflag2}
   \end{figure*}

   \begin{figure*}
   \includegraphics[width=18cm]{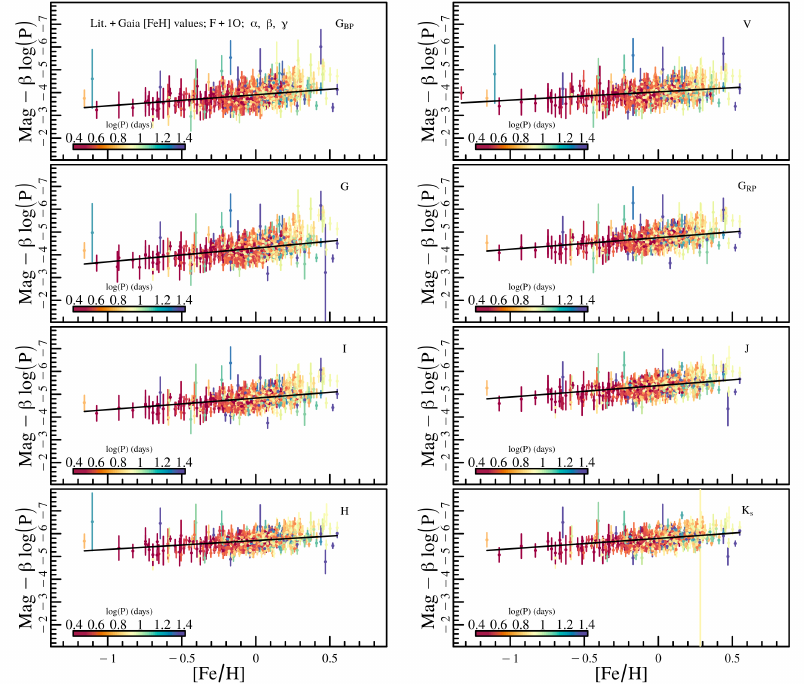}
   \caption{Visualization of the $PL$ dependence on the metallicity. The y-axis is the magnitudes subtracted by the period contribution and  the x-axis is the metallicity. The colour bar represents the logarithm of the period.}
              \label{fig:AbsMagLogP_FeH_abc_F1O_sflag2}
   \end{figure*}

   \begin{figure}
   \includegraphics[width=9cm]{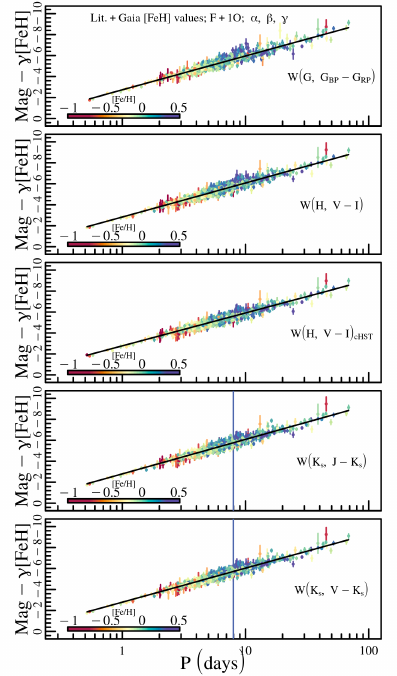}
   \caption{Same as Fig.~\ref{fig:AbsMagFeH_logP_abc_F1O_sflag2} but for the $PWZ$ relations.}
              \label{fig:AbsWesFeH_logP_abc_F1O_sflag2}
   \end{figure}

      \begin{figure}
   \includegraphics[width=9cm]{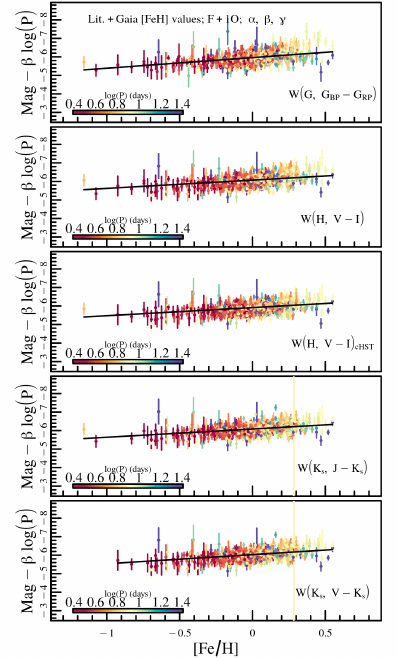}
   \caption{Same as Fig.~\ref{fig:AbsMagLogP_FeH_abc_F1O_sflag2} but for the $PWZ$ relations.}
              \label{fig:AbsWesLogP_FeH_abc_F1O_sflag2}
   \end{figure}

\subsection{Metallicity}

 Metallicities were taken from various literature sources. A large part of the sample is the same as in \citet[][see their sect. 4.1]{Trentin2023}. More specifically, we considered the large compilation of homogenised literature iron abundances for 436 DCEPs presented by \citet[][G18 hereinafter]{Groenewegen2018}, complemented with literature results for a few additional stars by \citet[][GC17 hereinafter]{Gaia2017}. To these data, we added the following samples: (i) 49 DCEPs presented in \citet[][collectively called R21 hereinafter]{Catanzaro2020,Ripepi2021a,Ripepi2021b}; (ii) the sample of 65 DCEPs published in \citet[][called T23 hereafter]{Trentin2023}; (iii) the 104 stars by \citet[][K22 hereinafter]{Kovtyukh2022} after removing the large overlap with the R21 and T23 sample \citep[see][for full details and the homogenisation procedure we adopted in merging the samples]{Trentin2023}; and (iv) a few stars with HiRes metallicities from the GALAH Survey \citep[GALactic Archaeology with HERMES;][stars OGLE-GD-CEP-0059, OGLE-GD-CEP-0058, V1253 Cen]{Galah2021} and from the PASTEL catalogue \citep[][star OGLE-GD-CEP-0117]{Pastel2016}. 
With these additions, the total number of DCEPs with metallicities from HiRes spectroscopy is 635. Concerning homogeneity, apart from the few stars from GALAH, PASTEL, or GC17, the main samples adopted in this work are the G18, K22, and combined C-MetaLL data. As mentioned above, the G18 sample is already homogeneous, while the K22 abundances were homogenised with those of the C-MetaLL sample in \citet{Trentin2023}. As discussed in \citet{Ripepi2021b}, the C-MetaLL data have only two stars in common with G18, namely X Sct and V5567 Sgr. For these two stars, the abundances agree well within 0.5 $\sigma$.

In addition to this already large sample, we decided to exploit the recent results released by the \gaia\ DR3. To this aim, we cross-matched the DCEP catalogue published by \gaia\ DR3 (Vizier catalogue \textit{I/358/vcep}) as amended by \citep{Ripepi2022a} with that of the astrophysical parameters (Vizier catalogue \textit{I/355/paramp}), retaining only matching stars that have a global metallicity value ([M/H]) derived from the medium-resolution spectra obtained with the Radial Velocity Spectrometer \citep[RVS, see][for details]{RecioBlanco2023}. The total number of matching stars is 983, among which 475  have HiRes metallicity estimations from the literature.
\gaia\ metallicities were corrected following the recipe by \citet[][their eq. 4 and table 5]{RecioBlanco2023}, and then we assigned a data-quality flag to every object.  Assigned flags are 1, 2, and 3, which refer to high-, medium-, and low-quality, respectively, according to 
\citet[][see sect. 9]{RecioBlanco2023}. 
 The Gaia abundances were derived using the stacked RVS spectra over all the epochs of observations. The resulting spectra are therefore an average of many spectra over the pulsation cycle. As the DCEPs vary in terms of effective temperature, surface gravity, and microturbulent velocity along the pulsation cycle, we can in principle expect an impact on the derived abundances. In addition, the abundances published in the Gaia Astrophysical parameters are the mean metallicities [M/H], which may differ from the [Fe/H] scale used for the HiRes sample. To investigate these potential concerns, we compared the corrected \gaia\ metallicities with those from the HiRes sample for the 475 stars in common. The result is shown in Fig.~\ref{fig:metallicityComparison}. There is good overall agreement for all the \gaia\ subsamples (low, medium, and high quality). In all cases, the average difference between HiRes and \gaia\ results is very low, namely $\Delta \sim -0.03$ dex, with a comfortable moderate dispersion of the order of 0.11-0.13 dex (worse for the low-quality \gaia\ data), indicating that the \gaia\ spectroscopic results are of comparable quality to the HiRes ones  overall and that the data can be used together. This is especially true for [Fe/H]$_{HiRes} > -0.3$ dex. Below this value, the \gaia\ sample is dominated by low-quality data and the dispersion becomes significantly larger (top panel). 
We corrected for the small offset of the \gaia\ data and decided to use only the medium- and high-quality data  in
the following in order to avoid including some unreliable abundance values from the low-quality data. The resulting sample is then composed of 910 DCEPs divided into 282 DCEP\_1Os, 22 DCEP\_1O2Os, 581 DCEP\_Fs, and 25 DCEP\_F1Os, where DCEP\_1O2Os and DCEP\_F1Os represent multi-mode pulsators. For these two last cases, we adopted the longest period, that is, the 1O period for the 1O2O pulsators and the F one for the F1O DCEPs. Therefore, our sample includes the equivalent of 304 1O and 606 F mode DCEPs.  

\subsection{Astrometry}\label{astrometry}

To carry out our analysis, we adopted the parallaxes from \gaia\ DR3, which were individually corrected for the zero point offset, adopting the recipe by \citet[][L21]{Lindegren2021}\footnote{\url{https://www.cosmos.esa.int/web/gaia/edr3-code} }.  We highlight the characteristic shape of the correction, with a sharp hump around $G\sim12$ mag, the explanation for which can be found in \citet{Lindegren2021}. 
The individual corrections are shown in Fig.~\ref{fig:parallaxCorrections} as a function of the magnitude and of the ecliptic latitude. 
As criteria of the goodness of the astrometry, we adopted two indicators: (i) the {\tt fidelity\_v2} index as tabulated by \citet{Rybizki2022}, retaining only objects with values larger than 0.5, and (ii) the {\tt RUWE} parameter published in \gaia\ DR3. Although the \gaia\ documentation recommends using a threshold below 1.4 for good astrometric solutions, we choose to be less conservative and use a slightly larger threshold equal to 1.5. This choice allows us to retain possibly good objects falling just outside the 1.4 threshold. In any case, our robust outlier-removal procedure (see Sect.~\ref{sec:method}) removes possible deviating stars introduced by the slightly enlarged adopted threshold.  

The simultaneous use of the {\tt fidelity\_v2} and {\tt RUWE} parameters led us to reject 78 objects from our sample. A large fraction of the rejected DCEPs, namely 35, have $G \leq 7$ mag. This is not surprising, as  \gaia\ parallaxes are known to be uncertain for such bright stars \citep[e.g.][]{Lindegren2021}.

\section{Derivation of the $PLZ/PWZ$ relations} \label{sec:method}

In this section, we describe the procedure adopted to calibrate the $PLZ/PWZ$ relations. 
 The approach is the same as in \citet{Ripepi2020} and \citet{Ripepi2021a}. 
To avoid any bias, the whole DCEPs sample was considered, including negative parallaxes, and without any selection on the parallax relative errors. Moreover, we adopted the astrometry-based-luminosity (ABL) formalism \citep{Feast1997, Arenou1999}, which allows us to treat the parallax $\varpi$ as a linear parameter:

\begin{equation}
    {\rm ABL}=\varpi 10^{0.2 m -2}=10^{0.2(\alpha+(\beta+\delta {\rm [Fe/H])}(\log P-\log P_0) +\gamma {\rm [Fe/H]})}, \label{eq:abl}  
\end{equation}

\noindent 
where the parallax $\varpi$, the apparent generic magnitude $m$,
 the period $P,$ and the metallicity $[Fe/H]$ are the observables, while the unknowns are the four parameters $\alpha,\beta,\gamma$, and $\delta$ (intercept, slope, metallicity dependence of the intercept, and the  metallicity dependence of the slope, respectively). The $P_{0}$ quantity is a pivoting period ($10 d$) adopted to reduce the correlation between the $\alpha$ and $\beta$ parameters, which dominate the $PL$ and $PW$ relations.

To take into account the presence of possible outlier measurements, we applied multiple sigma-clipping removals but limited the number of rejected data to $\sim 10\%$ of the full sample.
Based on the results described in the previous papers of this series, we decided to exclude the case with no metallicity dependence at all (i.e. $\gamma \neq 0$) in Eq.~\ref{eq:abl}. 
Finally, the 1O pulsators are included in the fitting sample by fundamentalising their periods according to the relation by \citet{Alcock1995} and \citet{Feast1997}.
In the following sections, we consider two data samples: (i) the \textit{Lit+DR3} sample, which includes all the selected sources introduced in Sect.~\ref{sec:data}; and (ii) the \textit{Lit} sample, which was obtained by excluding the sources with only a Gaia DR3 metallicity estimate, therefore retaining only DCEPs with metallicity measures from ground-based HiRes spectrographs. 

The uncertainties on the parameters calculated in the following analysis are obtained through the bootstrap technique. A set of 1000 resampling experiments is performed and the coefficients of the Eq.~\ref{eq:abl} are calculated for each obtained random data set. Finally, the quoted errors are estimated by considering the robust standard deviation of the obtained coefficient distributions.


\begin{table*}
\footnotesize\setlength{\tabcolsep}{3pt}
\centering
\caption{Fitting results from the Lit. data set.} 
\label{tab:fitResSflag0}
\begin{tabular}{cccccccccccccc}
  \hline
ID & $\alpha$ & $\beta$ & $\gamma$ & $\delta$ & $\mu_0^{LMC}$ & $rms$ & Band & Mode & $N_{dat}$ \\ 
  \hline
 1 & $-4.253 \pm 0.026$ & $-2.591 \pm 0.061$ & $-0.403 \pm 0.133$ & $        -       $ & $18.484 \pm 0.064$ & 0.020 &  $PLG$                  & F    & 386 \\ 
 2 & $-4.258 \pm 0.029$ & $-2.610 \pm 0.065$ & $-0.370 \pm 0.143$ & $ 0.211 \pm 0.320$ & $18.453 \pm 0.067$ & 0.020 &  $PLG$                  & F    & 388 \\ 
 3 & $-4.268 \pm 0.025$ & $-2.529 \pm 0.057$ & $-0.431 \pm 0.108$ & $        -       $ & $18.485 \pm 0.049$ & 0.021 &  $PLG$                  & F+1O & 503 \\ 
 4 & $-4.275 \pm 0.026$ & $-2.551 \pm 0.057$ & $-0.367 \pm 0.129$ & $ 0.272 \pm 0.207$ & $18.447 \pm 0.042$ & 0.022 &  $PLG$                  & F+1O & 510 \\ 
 5 & $-3.843 \pm 0.020$ & $-2.361 \pm 0.060$ & $-0.417 \pm 0.112$ & $        -       $ & $18.423 \pm 0.052$ & 0.022 &  $PLG_{BP}$             & F    & 377 \\ 
 6 & $-3.845 \pm 0.022$ & $-2.367 \pm 0.064$ & $-0.407 \pm 0.124$ & $ 0.070 \pm 0.316$ & $18.414 \pm 0.060$ & 0.022 &  $PLG_{BP}$             & F    & 377 \\ 
 7 & $-3.897 \pm 0.028$ & $-2.286 \pm 0.062$ & $-0.330 \pm 0.123$ & $        -       $ & $18.498 \pm 0.058$ & 0.025 &  $PLG_{BP}$             & F+1O & 498 \\ 
 8 & $-3.905 \pm 0.030$ & $-2.311 \pm 0.065$ & $-0.255 \pm 0.148$ & $ 0.335 \pm 0.244$ & $18.451 \pm 0.050$ & 0.024 &  $PLG_{BP}$             & F+1O & 500 \\ 
 9 & $-4.702 \pm 0.018$ & $-2.667 \pm 0.054$ & $-0.481 \pm 0.094$ & $        -       $ & $18.384 \pm 0.044$ & 0.012 &  $PLG_{RP}$             & F    & 376 \\ 
10 & $-4.706 \pm 0.019$ & $-2.681 \pm 0.057$ & $-0.461 \pm 0.106$ & $ 0.162 \pm 0.310$ & $18.356 \pm 0.056$ & 0.012 &  $PLG_{RP}$             & F    & 373 \\ 
11 & $-4.725 \pm 0.018$ & $-2.606 \pm 0.051$ & $-0.460 \pm 0.083$ & $        -       $ & $18.420 \pm 0.037$ & 0.014 &  $PLG_{RP}$             & F+1O & 489 \\ 
12 & $-4.736 \pm 0.019$ & $-2.631 \pm 0.049$ & $-0.385 \pm 0.104$ & $ 0.232 \pm 0.180$ & $18.407 \pm 0.035$ & 0.014 &  $PLG_{RP}$             & F+1O & 494 \\ 
13 & $-5.669 \pm 0.016$ & $-3.103 \pm 0.046$ & $-0.374 \pm 0.081$ & $        -       $ & $18.441 \pm 0.035$ & 0.008 &  $PLH$                  & F    & 384 \\ 
14 & $-5.668 \pm 0.017$ & $-3.107 \pm 0.050$ & $-0.385 \pm 0.088$ & $ 0.045 \pm 0.270$ & $18.440 \pm 0.049$ & 0.008 &  $PLH$                  & F    & 383 \\ 
15 & $-5.684 \pm 0.014$ & $-3.068 \pm 0.042$ & $-0.386 \pm 0.070$ & $        -       $ & $18.454 \pm 0.028$ & 0.009 &  $PLH$                  & F+1O & 505 \\ 
16 & $-5.688 \pm 0.015$ & $-3.080 \pm 0.040$ & $-0.352 \pm 0.076$ & $ 0.161 \pm 0.124$ & $18.438 \pm 0.028$ & 0.009 &  $PLH$                  & F+1O & 508 \\ 
17 & $-4.781 \pm 0.017$ & $-2.723 \pm 0.050$ & $-0.444 \pm 0.089$ & $        -       $ & $18.411 \pm 0.042$ & 0.011 &  $PLI$                  & F    & 362 \\ 
18 & $-4.784 \pm 0.018$ & $-2.744 \pm 0.056$ & $-0.420 \pm 0.100$ & $ 0.285 \pm 0.322$ & $18.365 \pm 0.062$ & 0.011 &  $PLI$                  & F    & 359 \\ 
19 & $-4.805 \pm 0.017$ & $-2.632 \pm 0.049$ & $-0.472 \pm 0.081$ & $        -       $ & $18.439 \pm 0.036$ & 0.014 &  $PLI$                  & F+1O & 495 \\ 
20 & $-4.830 \pm 0.023$ & $-2.675 \pm 0.051$ & $-0.348 \pm 0.110$ & $ 0.355 \pm 0.176$ & $18.421 \pm 0.034$ & 0.014 &  $PLI$                  & F+1O & 504 \\ 
21 & $-5.346 \pm 0.017$ & $-2.948 \pm 0.054$ & $-0.506 \pm 0.093$ & $        -       $ & $18.397 \pm 0.042$ & 0.010 &  $PLJ$                  & F    & 391 \\ 
22 & $-5.345 \pm 0.018$ & $-2.947 \pm 0.055$ & $-0.509 \pm 0.102$ & $-0.014 \pm 0.304$ & $18.400 \pm 0.057$ & 0.010 &  $PLJ$                  & F    & 390 \\ 
23 & $-5.366 \pm 0.016$ & $-2.873 \pm 0.047$ & $-0.504 \pm 0.082$ & $        -       $ & $18.430 \pm 0.034$ & 0.012 & $PLJ$                   & F+1O & 520 \\ 
24 & $-5.369 \pm 0.019$ & $-2.878 \pm 0.047$ & $-0.479 \pm 0.090$ & $ 0.088 \pm 0.159$ & $18.427 \pm 0.037$ & 0.012 & $PLJ$                   & F+1O & 516 \\ 
25 & $-5.775 \pm 0.017$ & $-3.158 \pm 0.045$ & $-0.378 \pm 0.076$ & $        -       $ & $18.458 \pm 0.036$ & 0.007 &  $PLK$                  & F    & 368 \\ 
26 & $-5.774 \pm 0.018$ & $-3.176 \pm 0.051$ & $-0.386 \pm 0.089$ & $ 0.171 \pm 0.265$ & $18.426 \pm 0.048$ & 0.008 & $PLK$                   & F    & 376 \\ 
27 & $-5.780 \pm 0.015$ & $-3.121 \pm 0.037$ & $-0.433 \pm 0.065$ & $        -       $ & $18.444 \pm 0.029$ & 0.008 &  $PLK$                  & F+1O & 483 \\ 
28 & $-5.786 \pm 0.015$ & $-3.130 \pm 0.036$ & $-0.374 \pm 0.075$ & $ 0.191 \pm 0.114$ & $18.419 \pm 0.026$ & 0.008 &  $PLK$                  & F+1O & 487 \\ 
29 & $-3.974 \pm 0.019$ & $-2.372 \pm 0.058$ & $-0.370 \pm 0.103$ & $        -       $ & $18.432 \pm 0.051$ & 0.019 &  $PLV$                  & F    & 366 \\ 
30 & $-3.977 \pm 0.020$ & $-2.386 \pm 0.060$ & $-0.348 \pm 0.119$ & $ 0.108 \pm 0.337$ & $18.415 \pm 0.061$ & 0.019 &  $PLV$                  & F    & 365 \\ 
31 & $-4.002 \pm 0.020$ & $-2.299 \pm 0.060$ & $-0.396 \pm 0.103$ & $        -       $ & $18.453 \pm 0.049$ & 0.022 &  $PLV$                  & F+1O & 491 \\ 
32 & $-4.040 \pm 0.029$ & $-2.370 \pm 0.066$ & $-0.197 \pm 0.147$ & $ 0.482 \pm 0.247$ & $18.434 \pm 0.050$ & 0.023 &  $PLV$                  & F+1O & 501 \\ 
33 & $-5.910 \pm 0.017$ & $-3.234 \pm 0.054$ & $-0.637 \pm 0.085$ & $        -       $ & $18.380 \pm 0.037$ & 0.007 &  $PW_{G,G_{BP}-G_{RP}}$ & F    & 378 \\ 
34 & $-5.917 \pm 0.019$ & $-3.264 \pm 0.057$ & $-0.608 \pm 0.102$ & $ 0.436 \pm 0.318$ & $18.312 \pm 0.053$ & 0.007 &  $PW_{G,G_{BP}-G_{RP}}$ & F    & 385 \\ 
35 & $-5.925 \pm 0.016$ & $-3.199 \pm 0.043$ & $-0.553 \pm 0.072$ & $        -       $ & $18.424 \pm 0.030$ & 0.008 &  $PW_{G,G_{BP}-G_{RP}}$ & F+1O & 499 \\ 
36 & $-5.924 \pm 0.016$ & $-3.198 \pm 0.043$ & $-0.566 \pm 0.092$ & $-0.042 \pm 0.183$ & $18.422 \pm 0.032$ & 0.008 &  $PW_{G,G_{BP}-G_{RP}}$ & F+1O & 499 \\ 
37 & $-6.045 \pm 0.016$ & $-3.253 \pm 0.050$ & $-0.426 \pm 0.085$ & $        -       $ & $18.442 \pm 0.035$ & 0.007 &  $PW_{H,V-I}$           & F    & 392 \\ 
38 & $-6.048 \pm 0.017$ & $-3.273 \pm 0.051$ & $-0.409 \pm 0.088$ & $ 0.207 \pm 0.293$ & $18.410 \pm 0.056$ & 0.007 &  $PW_{H,V-I}$           & F    & 394 \\ 
39 & $-6.055 \pm 0.015$ & $-3.242 \pm 0.040$ & $-0.454 \pm 0.069$ & $        -       $ & $18.434 \pm 0.027$ & 0.008 &  $PW_{H,V-I}$           & F+1O & 516 \\ 
40 & $-6.059 \pm 0.015$ & $-3.248 \pm 0.039$ & $-0.404 \pm 0.077$ & $ 0.157 \pm 0.128$ & $18.426 \pm 0.028$ & 0.008 &  $PW_{H,V-I}$           & F+1O & 512 \\ 
41 & $-5.908 \pm 0.014$ & $-3.167 \pm 0.040$ & $-0.338 \pm 0.067$ & $        -       $ & $18.421 \pm 0.032$ & 0.006 &  $PW_{H,V-I}^{cHST}$    & F    & 369 \\ 
42 & $-5.903 \pm 0.015$ & $-3.207 \pm 0.049$ & $-0.366 \pm 0.078$ & $ 0.366 \pm 0.259$ & $18.325 \pm 0.047$ & 0.007 &  $PW_{H,V-I}^{cHST}$    & F    & 393 \\ 
43 & $-5.915 \pm 0.013$ & $-3.133 \pm 0.036$ & $-0.369 \pm 0.063$ & $        -       $ & $18.414 \pm 0.028$ & 0.008 &  $PW_{H,V-I}^{cHST}$    & F+1O & 491 \\ 
44 & $-5.922 \pm 0.013$ & $-3.156 \pm 0.035$ & $-0.322 \pm 0.071$ & $ 0.267 \pm 0.110$ & $18.369 \pm 0.026$ & 0.008 &  $PW_{H,V-I}^{cHST}$    & F+1O & 499 \\ 
45 & $-6.066 \pm 0.015$ & $-3.297 \pm 0.041$ & $-0.414 \pm 0.072$ & $        -       $ & $18.433 \pm 0.034$ & 0.007 &  $PW_{K_S,J-K_S}$       & F    & 381 \\ 
46 & $-6.066 \pm 0.017$ & $-3.345 \pm 0.054$ & $-0.389 \pm 0.084$ & $ 0.598 \pm 0.285$ & $18.297 \pm 0.056$ & 0.007 &  $PW_{K_S,J-K_S}$       & F    & 393 \\ 
47 & $-6.069 \pm 0.015$ & $-3.260 \pm 0.040$ & $-0.440 \pm 0.061$ & $        -       $ & $18.409 \pm 0.027$ & 0.007 &  $PW_{K_S,J-K_S}$       & F+1O & 500 \\ 
48 & $-6.067 \pm 0.014$ & $-3.255 \pm 0.035$ & $-0.375 \pm 0.070$ & $ 0.203 \pm 0.112$ & $18.396 \pm 0.025$ & 0.007 &  $PW_{K_S,J-K_S}$       & F+1O & 498 \\ 
49 & $-6.023 \pm 0.016$ & $-3.267 \pm 0.047$ & $-0.383 \pm 0.077$ & $        -       $ & $18.447 \pm 0.035$ & 0.007 &  $PW_{K_S,V-K_S}$       & F    & 381 \\ 
50 & $-6.027 \pm 0.018$ & $-3.276 \pm 0.052$ & $-0.382 \pm 0.088$ & $ 0.155 \pm 0.270$ & $18.433 \pm 0.049$ & 0.007 &  $PW_{K_S,V-K_S}$       & F    & 380 \\ 
51 & $-6.022 \pm 0.015$ & $-3.228 \pm 0.037$ & $-0.464 \pm 0.064$ & $        -       $ & $18.417 \pm 0.028$ & 0.007 &  $PW_{K_S,V-K_S}$       & F+1O & 503 \\ 
52 & $-6.026 \pm 0.015$ & $-3.234 \pm 0.037$ & $-0.414 \pm 0.075$ & $ 0.169 \pm 0.118$ & $18.402 \pm 0.027$ & 0.008 &  $PW_{K_S,V-K_S}$       & F+1O & 506 \\ 
   \hline
\end{tabular}
\tablefoot{The meaning of the columns is the same as in Table \ref{tab:fitResSflag2}.}
\end{table*}


   \begin{figure}
   \includegraphics[width=9cm]{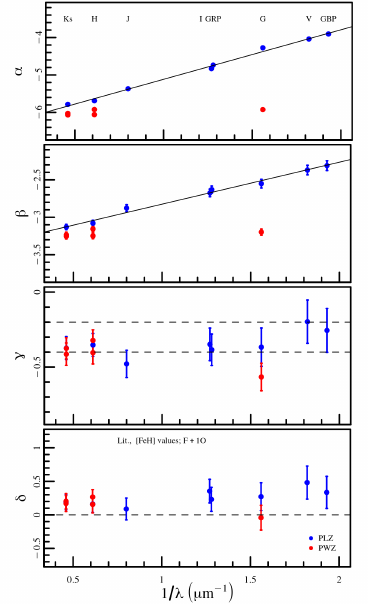}
   \caption{Same as Fig.~\ref{fig:coeffs-vs-wlen-abcd-F1O-sflag2} but for the Lit. sample.}
              \label{fig:coeffs-vs-wlen-abcd-F1O-sflag0}
   \end{figure}

      \begin{figure}
   \includegraphics[width=9cm]{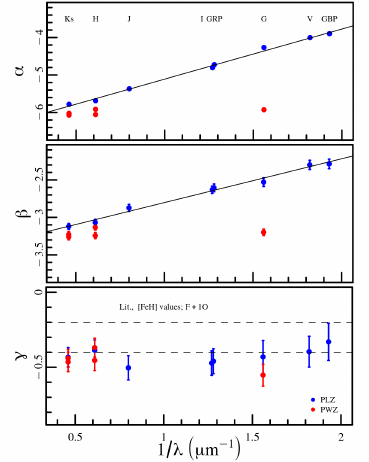}
   \caption{Same as Fig.~\ref{fig:coeffs-vs-wlen-abc-F1O-sflag2} but for the Lit. sample.}
              \label{fig:coeffs-vs-wlen-abc-F1O-sflag0}
   \end{figure}

\section{Results}\label{sec:results}

\subsection{Literature and \gaia\ DR3 sample}

The results of the fitting procedure obtained for this case are listed in Table ~\ref{tab:fitResSflag2}. We studied the dependence of the fitted $PLZ/PWZ$ coefficients on the central wavelength of the considered filters. To associate a characteristic wavelength to the Wesenheit magnitudes, we decided to consider the central wavelength of the main band (e.g. $G$ band for the $W_{G, BP-RP}$ case). 
The results of this analysis are shown in Fig.~\ref{fig:coeffs-vs-wlen-abcd-F1O-sflag2} for the F+1O sample (i.e. the sample including both F and fundamentalised 1O pulsators) and assuming a metallicity dependence also in the slope of the $PLZ/PWZ$ relation (i.e. $\delta \neq 0$).
Looking at the $PLZ$ coefficients (blue filled circles), a strong linear dependence as a function of the $\lambda^{-1}$ is evident only for the $\alpha$ and $\beta$ coefficients (correlation coefficient $R^2\simeq 1$), while for $\gamma$ and $\delta$ the slope of the fit is consistent with zero. The relations between $\alpha$ , $\beta,$ and $\lambda^{-1}$ are the following:
\begin{equation*}
    \alpha = (0.618\pm0.003)\cdot \lambda^{-1} + (-3.428\pm0.006),
\end{equation*}

\noindent
with  $rms=0.019, R^2=0.99$, and

\begin{equation*}
     \beta = (1.345\pm0.026)\cdot\lambda^{-1} + (-6.49\pm0.04), 
\end{equation*}

\noindent
with  $rms=0.059, R^2=0.99$. The strong dependence of the slope on the wavelength is a well-known feature, as thoroughly discussed by \citet{Madore2011}.

A similar discussion is valid for the case where the $\delta$ coefficient is neglected in  Eq.~\ref{eq:abl}. The estimated parameters are shown in Fig.~\ref{fig:coeffs-vs-wlen-abc-F1O-sflag2}, while the linear equations for $\alpha$ and $\beta$ coefficients are the following:

\begin{equation*}
    \alpha = (0.64\pm0.04)\cdot \lambda^{-1} + (-3.45\pm0.03),
\end{equation*}
\noindent 
with  $rms=0.02, R^2=0.99$, and

\begin{equation*}
     \beta = (1.37\pm0.04)\cdot\lambda^{-1} + (-6.50\pm0.05), 
\end{equation*}
\noindent
with  $rms=0.06, R^2=0.99$.

Comparing these results with those in the previous case, we found good agreement between all the parameters. To easily compare our $\gamma$ values with those in the recent literature, we traced two reference lines in the third panel of Figs.~\ref{fig:coeffs-vs-wlen-abcd-F1O-sflag2} and~\ref{fig:coeffs-vs-wlen-abc-F1O-sflag2}  corresponding to  $\gamma = -0.2$ and $-0.4$ dex. Indeed, this range of values includes almost all the recent estimates for the value of $\gamma$ \citep[see Table 1 of][and Sect.~ \ref{sec:comparison} for a comprehensive list of recent results]{Breuval2022}. 
Concerning the $\delta$ parameter, the bottom panel of Fig.~\ref{fig:coeffs-vs-wlen-abcd-F1O-sflag2} shows a significant metallicity dependence of the slopes only in the cases of the $V$-band $PL$. We note that in this case, the corresponding $\gamma$ coefficients is highly consistent with the average literature values.

For the sake of completeness, in Appendix~\ref{appA} we also plot the results obtained by fitting the Eq.~\ref{eq:abl} to the sample including only the F pulsators (see Fig.~\ref{fig:coeffs_vs_wlen_abcd_F_sflag2} and Fig.~\ref{fig:coeffs_vs_wlen_abc_F_sflag2}). The exclusion of the fundamentalised 1O pulsators increases the errors on all the coefficients, especially for $\delta$. More specifically, we notice that the $\alpha$ and the $\beta$ coefficients,
within the errors, are found to be slightly increased and decreased, respectively, while the $\gamma$ values are bigger (in an absolute sense), except for cases of the $G$ and $\,G_{bp}$ magnitudes. The $\delta$ parameter assumes also higher values in general and is more scattered compared with the F+1O case, but the large uncertainties prevent us from drawing firm conclusions.

To help in visualisation of the fitted relations listed in Table~\ref{tab:fitResSflag2}, we considered $\delta=0$, because in this case, the dependence on $\log(P)$ in Eq.~\ref{eq:abl} can be separated from that on metallicity. Figures~\ref{fig:AbsMagFeH_logP_abc_F1O_sflag2} and~\ref{fig:AbsMagLogP_FeH_abc_F1O_sflag2} show the projections of the $PLZ$ relations in two dimensions using $P$ and [Fe/H] on the x-axis, respectively. Similarly, Figs.~\ref{fig:AbsWesFeH_logP_abc_F1O_sflag2} and~\ref{fig:AbsWesLogP_FeH_abc_F1O_sflag2} display the same projections for the $PWZ$ relations.
Figures~\ref{fig:AbsMagFeH_logP_abc_F1O_sflag2} and~\ref{fig:AbsWesFeH_logP_abc_F1O_sflag2} show no colour trend with metallicity, indicating that the correction for this parameter was effective. On the other hand, Figs.~\ref{fig:AbsMagLogP_FeH_abc_F1O_sflag2} and~\ref{fig:AbsWesLogP_FeH_abc_F1O_sflag2} display the metallicity dependence of the intercepts of the $PL$ and $PW$ relations, respectively,  in a direct way. In these figures, the slopes of the solid lines, representing the fits for the different magnitudes, are a direct visualisation of the values of $\gamma$. It is possible to appreciate the relevance and importance of the extension towards the metal-poor regime for the determination of this parameter. Indeed, even a small number of objects with [Fe/H]$<-0.4$ dex can  significantly affect and constrain the slope in these plots.

  \begin{figure}
   \includegraphics[width=9.0cm]{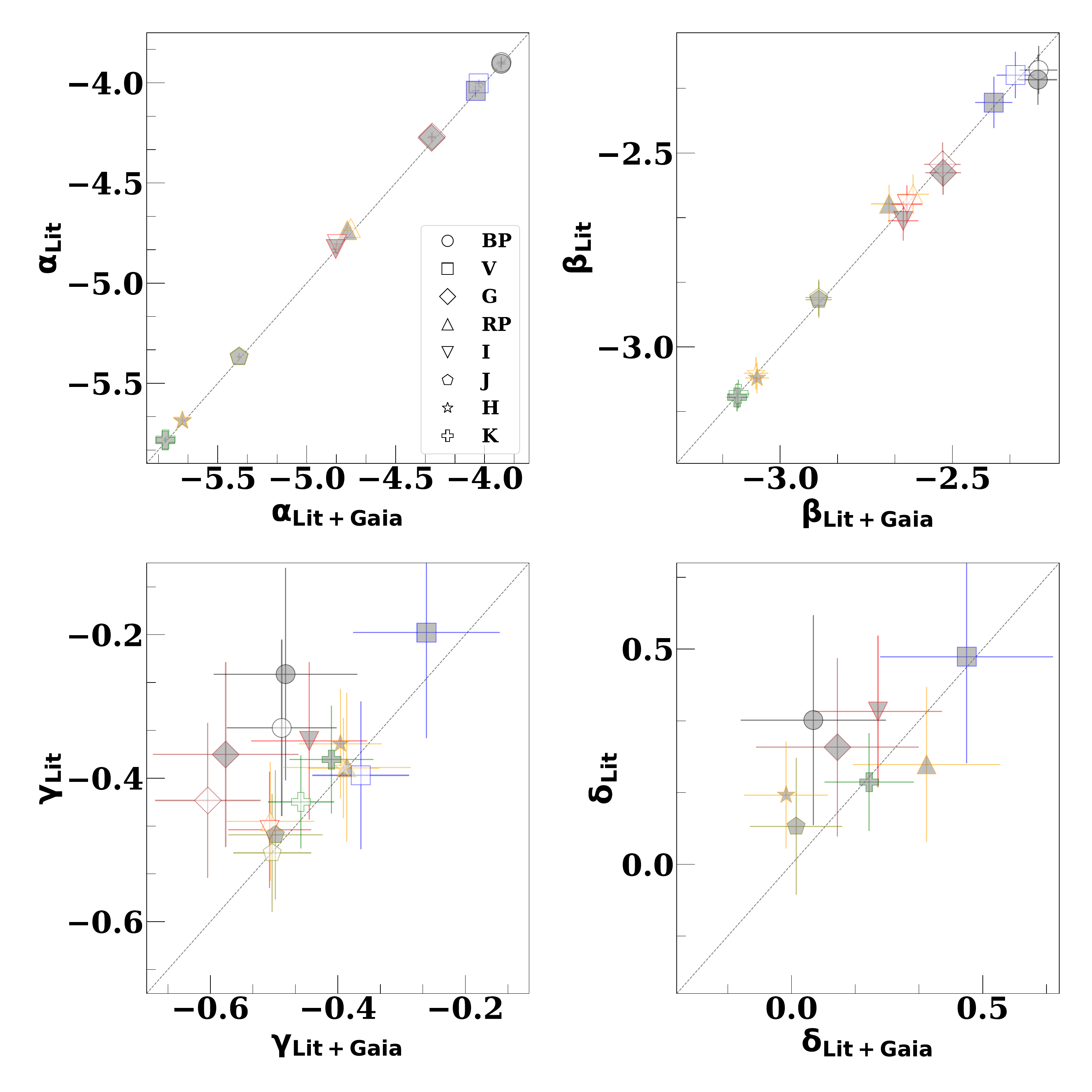}
   \caption{Comparison between the $PLZ$ coefficients obtained with the Lit.+Gaia data set and the Lit. data set. From top-left to bottom-right, the panels show the results for the $\alpha$, $\beta$, $\gamma,$ and $\delta$ coefficients, respectively. For each coefficient, different photometric bands are plotted using different symbols, as labelled in the top-left panel. Grey-filled and empty symbols indicate the fit results including or excluding the $\delta$ parameter, respectively.}
\label{fig:LitVsLitgaia.pl}
   \end{figure}

\begin{figure}
   \includegraphics[width=9.0cm]{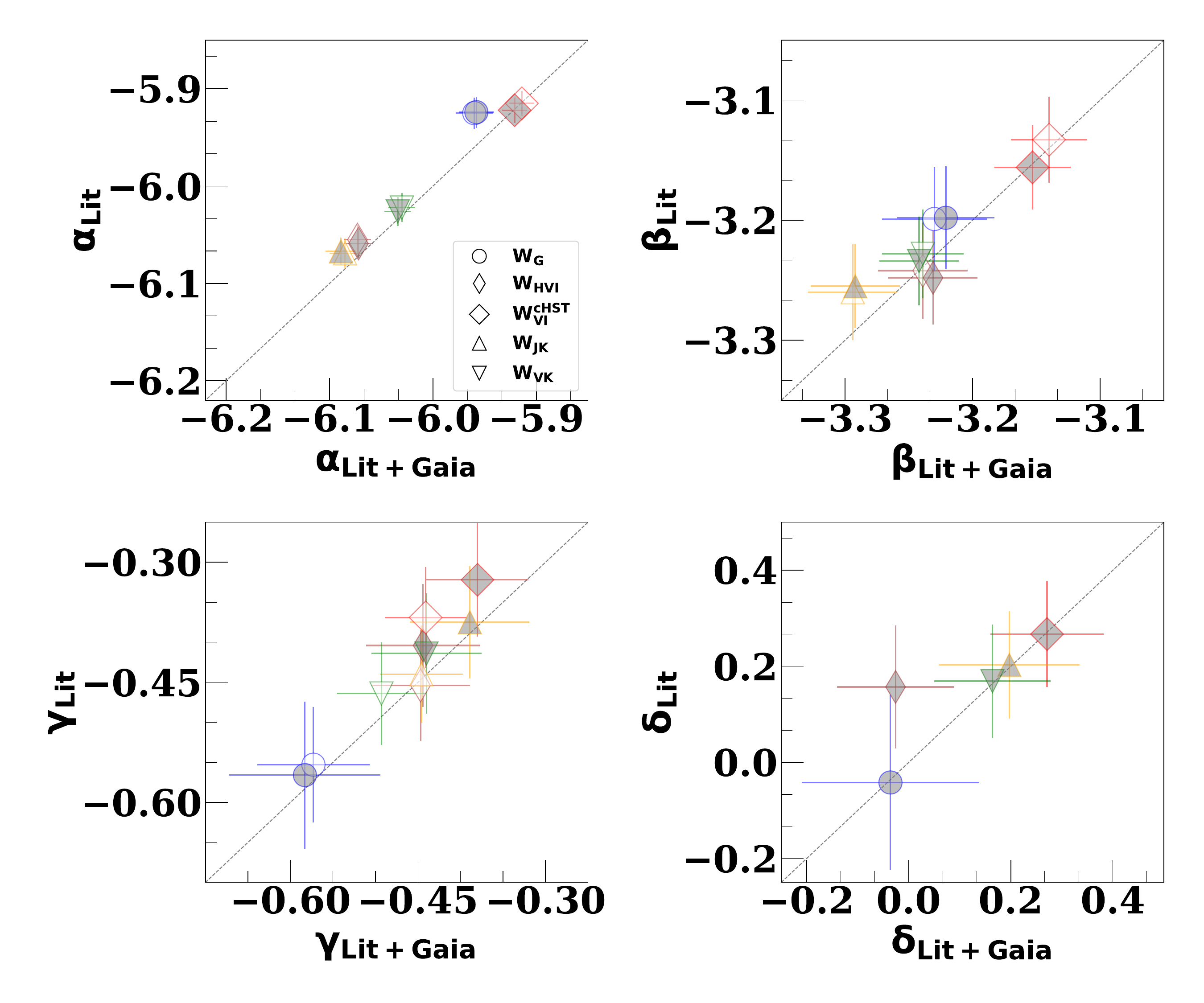}
   \caption{Same as Fig.~\ref{fig:LitVsLitgaia.pl} but for the $PWZ$ relations.}
    \label{fig:LitVsLitgaia.pw}
\end{figure}

\subsection{Literature sample only}
\label{sect:lit_sample}

The results of the fitting procedure obtained including the literature sample only are listed in  Table~\ref{tab:fitResSflag0}.
Similarly to the case above, we fitted $\alpha$ and $\beta$ coefficients as a function of $\lambda^{-1}$ of the specific band (only for $PLZ$). These fits are shown in the top two panels of Fig.~\ref{fig:coeffs-vs-wlen-abcd-F1O-sflag0} and their equations are:
\begin{equation*}
    \alpha = (0.563\pm0.017)\cdot \lambda^{-1} + (-3.37\pm 0.03),
\end{equation*}
\noindent
with  $rms=0.04, R^2=0.99$, and

\begin{equation*}
     \beta = (1.32\pm0.04)\cdot\lambda^{-1} + (-6.43\pm 0.06), 
\end{equation*} 
\noindent
with  $rms=0.06, R^2=0.99$.
The fits shown in Fig.~\ref{fig:coeffs-vs-wlen-abc-F1O-sflag2} 
($\delta=0$ case for 
the Lit. sample) for $\alpha$ and $\beta$ are respectively 

\begin{equation*}
    \alpha=(0.596\pm 0.016)\cdot \lambda^{-1} + (-3.382\pm 0.032),
\end{equation*} 
\noindent
with $rms=0.035, R^2=0.99$, and

\begin{equation*}
     \beta=(1.335\pm 0.037)\cdot \lambda^{-1} + (-6.443\pm0.051),
\end{equation*} 
\noindent
with $rms=0.057, R^2=0.99$.

The inclusion of the $\delta$ parameter in the fit of the Lit. sample only determines a general decrease (in an absolute sense) in the $\gamma$ values, which are closer to the -0.2,-0.4 mag/dex range typical of previous works. However, at the same time, the $\delta$ parameters show a considerable dependence of the $PLZ/PWZ$ slopes on metallicity. If instead the $\delta$ parameter is neglected, as shown in Fig.~\ref{fig:coeffs-vs-wlen-abc-F1O-sflag0}, we again
find an increase in the absolute value of $\gamma$, as already found for the Lit.+Gaia case.
Figures~\ref{fig:coeffs_vs_wlen_abcd_F_sflag0} and~\ref{fig:coeffs_vs_wlen_abc_F_sflag0} in Appendix~\ref{appA} report the above analysis considering only the F pulsators. A comparison of the results 
 obtained for the F+1O and the F samples reveals no significant differences. 

\subsection{Comparison between Lit.+Gaia and Lit. samples}

It is interesting to study the impact of the inclusion of the \gaia\ DCEP sample together with the HiRes one. To this aim, we compared the estimated parameters for both cases in Figs.~\ref{fig:LitVsLitgaia.pl} and~\ref{fig:LitVsLitgaia.pw} for the $PLZ$ and $PWZ$ relations, respectively.
While the $\alpha$ and $\beta$ parameters seem to be quite robust, for $\gamma$ and $\delta$ the Lit. sample presents slightly smaller and higher values (in an absolute sense) compared with the Lit.+Gaia data set, respectively. However, in almost all the cases, the differences are insignificant within 1$\sigma$. For this reason, we expect to find similar trends whether we use the Lit.+Gaia or the Lit. sample. Therefore, unless specified, we refer in the following discussions to the Lit.+Gaia sample only. In Appendix~\ref{appB}, the reader can find analogues figures to those shown here.

\section{Discussion}\label{sec:discussion}

\subsection{Parallax correction}

\begin{figure}
   \includegraphics[width=9.0cm]{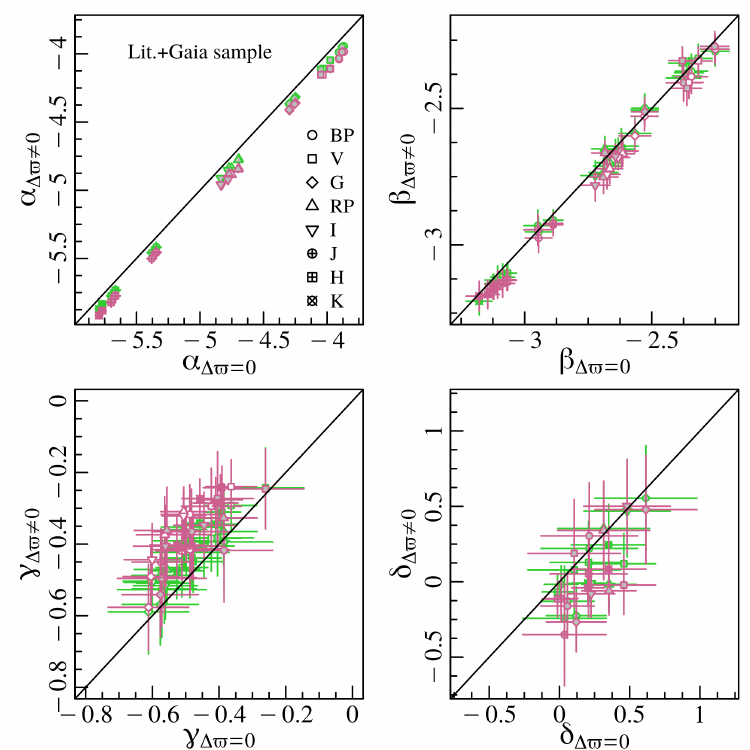}
   \caption{Comparison between the $PLZ$ coefficients obtained for different global offset values. The coefficient values obtained without offset assumption are chosen as reference and are on the x-axis, while those obtained whilst assuming the global parallax offset by \citet{Riess2021} and \citet{Molinaro2023} are on the y-axis and are plotted with green and pink symbols,  respectively.  From top-left to bottom-right, the panels show the results for the $\alpha$, $\beta$, $\gamma,$ and $\delta$ coefficients, respectively. For each coefficient, different photometric bands are plotted using different symbols, as labelled in the top-left panel, while grey-filled and white-filled symbols indicate the fit results including or excluding the $\delta$ parameter, respectively.}
\label{fig:OffsetComparisonLitGaia.pl}
   \end{figure}

\begin{figure}
   \includegraphics[width=9.0cm]{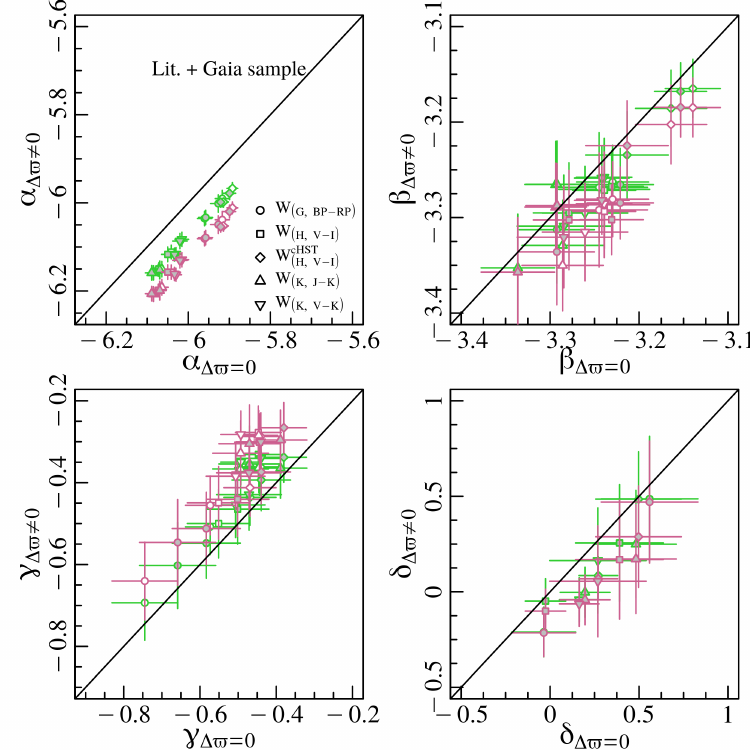}
   \caption{Same as Fig.~\ref{fig:OffsetComparisonLitGaia.pl} but for the $PWZ$ coefficients.}
\label{fig:OffsetComparisonLitGaia.pw}
   \end{figure}

A well-known feature of \gaia\ parallaxes in DR3 is the need for a global offset in the parallaxes after the L21 correction \citep[e.g.][and references therein]{Riess2021,Riess2022a,Molinaro2023}. The exact value of this offset is debated and appears to depend on the properties (e.g. location on the sky, magnitude, and colour) and kind of stellar tracer 
 adopted for its estimation \citep[see discussion in][]{Molinaro2023}. To take into account this feature, we also conducted the $PLZ/PWZ$ fit after adding the global offset to the EDR3 parallax values of $-14$ $\rm \mu$as \citep{Riess2021} or $-22$ $\rm \mu$as \citep{Molinaro2023}.
 A graphical comparison between these different cases for the Lit.+Gaia sample is shown in  Figs.~\ref{fig:OffsetComparisonLitGaia.pl} and ~\ref{fig:OffsetComparisonLitGaia.pw} for the $PLZ$ and $PWZ$, respectively. 
The introduction of the parallax correction increases the absolute values of $\alpha$ and $\beta$ by 1\%-4\%, while $\gamma$ decreases monotonically (in an absolute sense) for larger values of the parallax offset. This is particularly visible for the $PWZ$ relations, which show much less scatter than the $PLZ$ ones. In any case, the maximum variation is barely larger than 1$\sigma$. The $\delta$ coefficient varies less than $\gamma$ with a slightly decreasing trend; although this latter is insignificant at the 1$\sigma$ level. 
Overall, these effects have already been found and are coherent with those discussed in \citet{Ripepi2021a}.
In practice, the introduction of a significant global parallax zero point offset tends to reconcile our metallicity dependence with typical literature values, as $\gamma$ tends to diminish with increasing offset (both in an absolute sense). However, as discussed in the following section, the adoption of large offset values affects the absolute distance scale.

\subsection{Distance to the LMC}

\begin{figure}
   \includegraphics[width=9.0cm]{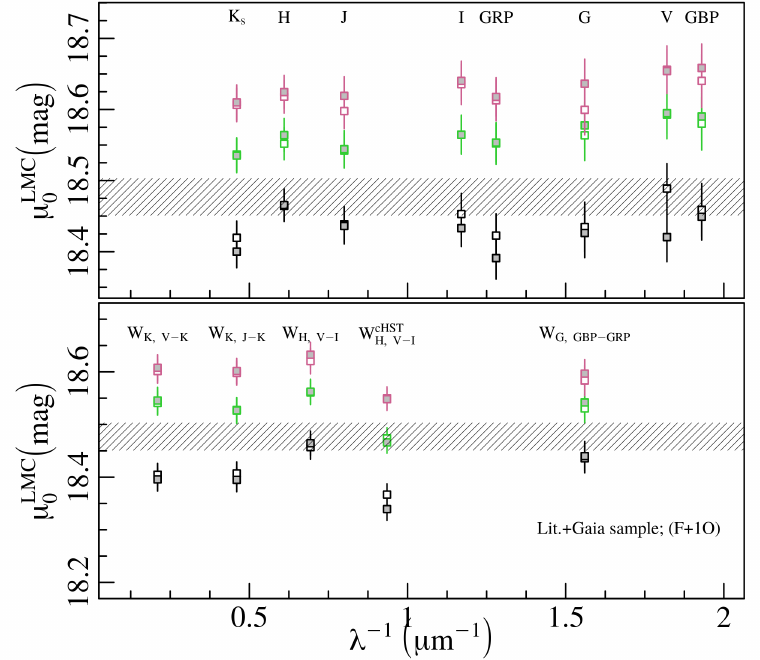}
   \caption{LMC distance modulus obtained using the relations of Table~\ref{tab:fitResSflag2} for the F+1O cases plotted against the central wavelength of the considered photometric bands. To avoid labels and symbols overlapping, the I band result has been shifted by -0.1$\mu m^{-1}$. The top panel shows the results for the $PLZ$ relations, while the bottom panel shows those for the $PWZ$ relations. Different colours indicate the results obtained for different parallax shift values: L21 correction (black squares), L21 + \citet{Riess2021} offset (green squares), L21 + \citet{Molinaro2023} (violet squares). Empty and grey-filled squares indicate the results obtained excluding and including the $\delta$ coefficient, respectively.}
\label{fig:OffsetComparisonLitGaia.lmcDistance}
   \end{figure}

To test the goodness of our $PLZ/PWZ$ relations and to understand which parameter affects the calibration, we used our relations to estimate the distance to the LMC and compared the results with the currently accepted geometric value $\mu_0=18.477\pm0.026$ mag \citep{Pietrzynski2019}. To this aim, we used the method described in detail in \citet{Ripepi2021a}. The distance moduli $\mu_0^{LMC}$ obtained in this way are listed in Tables~ \ref{tab:fitResSflag2} and \ref{tab:fitResSflag0} for the Lit.+Gaia and Lit. samples, respectively. The samples including only F pulsators generally give poorer results than the F+1O sample, and therefore in the following discussion we use only the latter data set. 
Figure~\ref{fig:OffsetComparisonLitGaia.lmcDistance} shows the resulting $\mu_0^{LMC}$ for the $PLZ/PWZ$ relations for Lit.+Gaia samples. The introduction of the global parallax correction worsens the distance estimation. 
More precisely, the results with no correction and with \citet{Riess2021} correction are very close to the lower and upper allowed limit, respectively. This suggests for the correction an intermediate value between the two cases, with the exception of $W_{H,V-I}^{cHST}$ where \citet{Riess2021} correction perfectly adjusts the value of $\mu_0^{LMC}$. 
Therefore, if we use the \citet{Pietrzynski2019} distance of the LMC as a reference, we have to conclude that the rather large values of $\gamma$ found in this work are favoured over the smaller ones from the recent literature.

\subsection{Metallicity sampling}

Although the results of our C-MetaLL project considerably enlarged the number of DCEPs with $[Fe/H]< -0.5$ dex, these objects represent less than 10\% of the pulsators used in this work, while the majority of the sources are distributed around the solar value. To test whether this unbalanced metallicity distribution could affect our determination of the $PLZ/PWZ$ relations, we conducted an experiment in order to recalculate these relationships by resampling the input data such that its metallicity distribution is uniform over the whole range. To this aim, we divided the sample into five bins in metallicity, imposing the same number of stars in each bin. 
 
\begin{figure}
   \includegraphics[width=9.0cm]{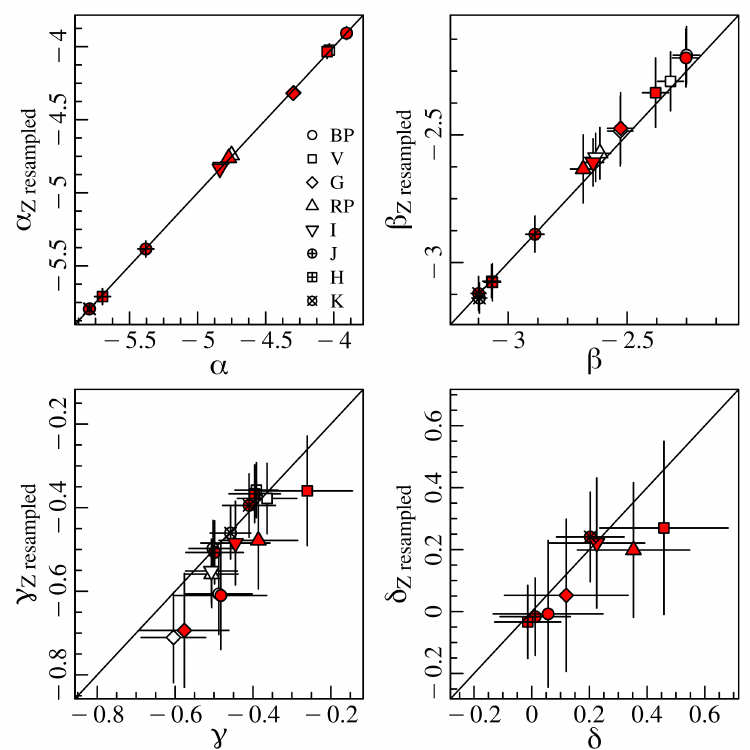}
   \caption{Comparison between the $PLZ$ coefficients obtained with the resampling in metallicity with those obtained using the entire data set. From top-left to bottom-right, the panels show the results for the $\alpha$, $\beta$, $\gamma,$ and $\delta$ coefficients, respectively. For each coefficient, different photometric bands are plotted using different symbols, as labelled in the top-left panel. In contrast, red-filled and white-filled symbols indicate the fit results including or excluding the $\delta$ parameter, 
respectively.}
\label{fig:samplingZ.pl}
   \end{figure}

\begin{figure}
   \includegraphics[width=9.0cm]{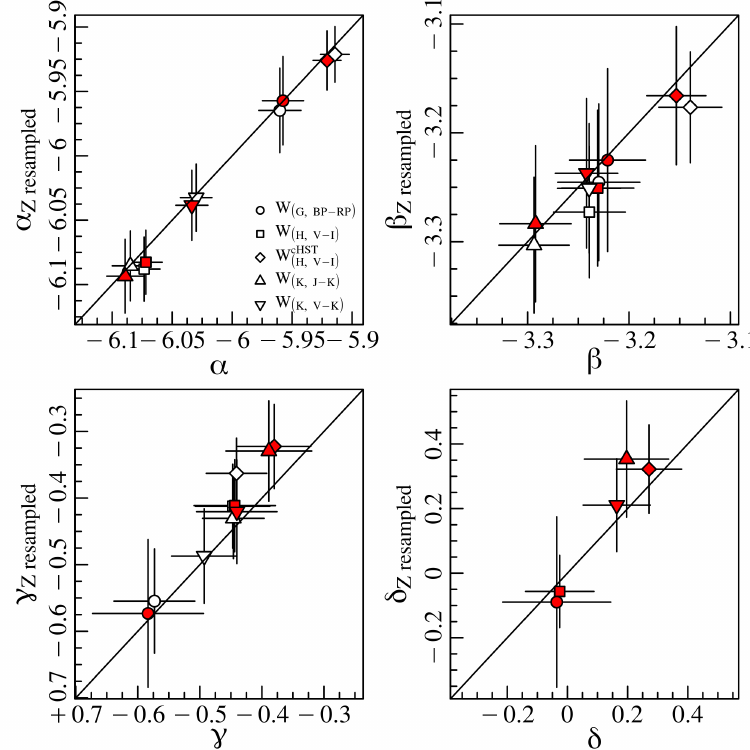}
   \caption{Same as Fig.~\ref{fig:samplingZ.2.pl}, but for the $PWZ$ coefficients.}
\label{fig:samplingZ.2.pl}
   \end{figure}

As there are few stars in the low-metallicity regime, the sources with $[Fe/H]<-0.5$ dex were all kept in the same bin and their number, $N_{[FeH<-0.5]}=44$ (considering the astrometric selection described in Sect.~\ref{astrometry}), set the number of pulsators included in each of the four remaining bins, whose size is approximately 0.25 dex in metallicity. The DCEPs populating each bin, apart from the lowest metallicity one, were randomly picked.  
Having resampled the data, we carried out the fit as in Sect.~\ref{sec:method} considering the Gaia+Lit. data set only, given that the other cases provide similar results. We also restricted the test to the F+1O sample, because the number of F pulsators in the most metal-poor bin is less than half of the F+1O combined sample, meaning that the total number of DCEPs used for the fit would be too small to achieve meaningful results. 
We repeated 10,000 times the fitting procedure, obtaining a distribution of values for the $\alpha,\,\beta,\,\gamma$, and $\delta$ coefficients. Their median values and relative uncertainties (scaled median absolute deviation (MAD)) are compared with the results obtained using the entire sample in  Fig.~\ref{fig:samplingZ.pl} and~\ref{fig:samplingZ.2.pl} for $PLZ$ and $PWZ$, respectively. 
Both figures show excellent agreement for all of the coefficients. Therefore, we can conclude that the results obtained in this work are not affected by the unbalanced metallicity distribution of the  sample. 

   \begin{figure*}
   \includegraphics[width=18cm]{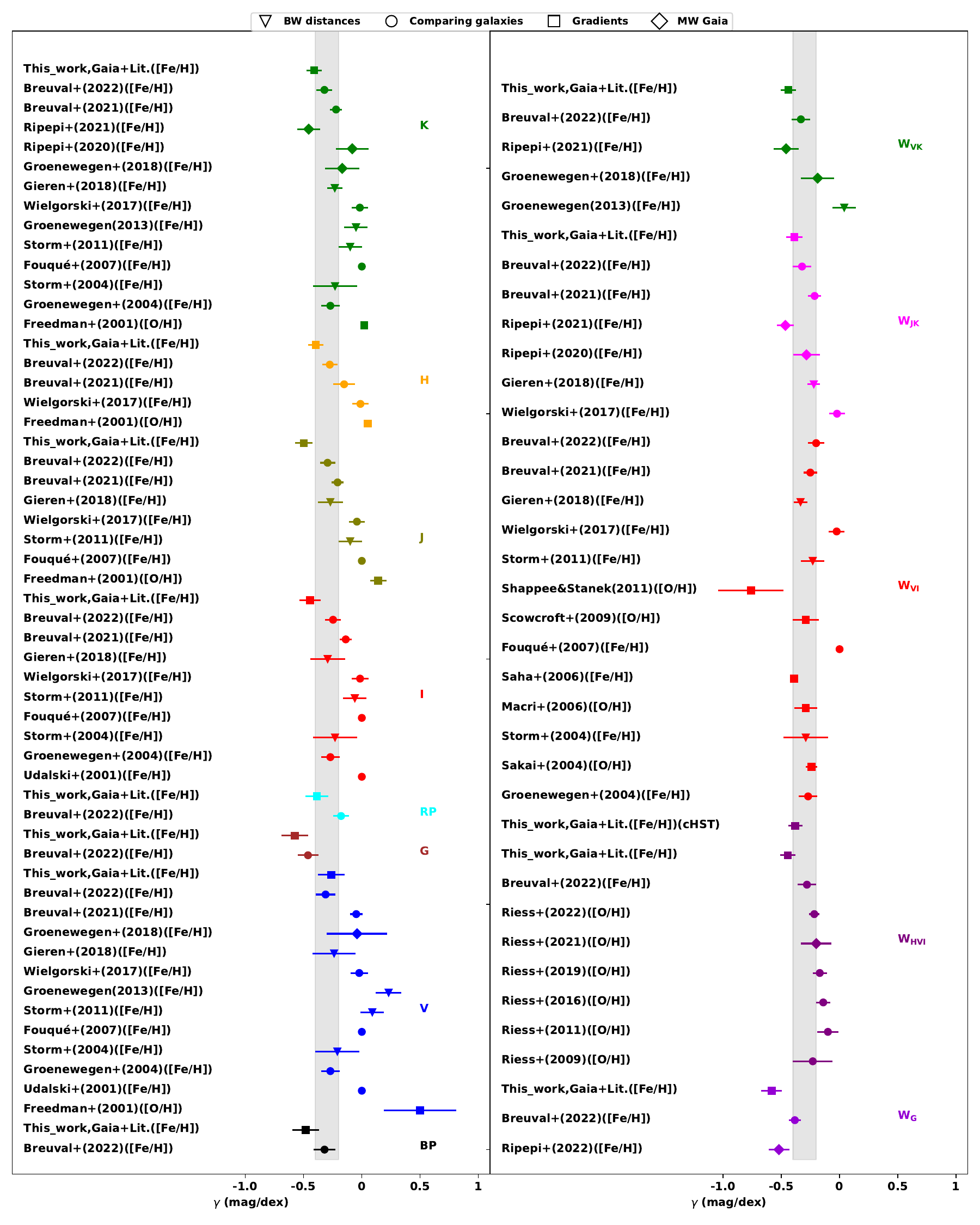}
   \caption{ Literature estimation of the $\gamma$ coefficient for $PLZ$ (left panel) and $PWZ$ (right panel) relations over the last 20 years. Bands are divided by colour and sorted by year of publication from bottom to top. Symbols correspond to the method used to derive the metallicity coefficient. The vertical shaded region corresponds to the range of intervals between $-0.4$ and $-0.2$ mag/dex (see text for detail). For the complete list of sources, see Fig.1 from \citet{Romaniello2008} and Table 1 from \cite{Breuval2022}.}
              \label{fig:literature_metallicity}
   \end{figure*}
   
\subsection{ Comparison with the literature}

\label{sec:comparison}

Figure~\ref{fig:literature_metallicity} shows several empirical estimations of $\gamma$ throughout the last 20 years. Results from the present work are also plotted, considering the case of the Lit.+Gaia data sample.
In more detail, different techniques have been used in order to analyse the metallicity effect on the $PL$ and $PW$ relations. Early studies adopting distances from the Baade-Wesselink (BW) analysis \citep[e.g.][]{Storm2004,Storm2011,groenewegen2013baade} reported very small values for the $\gamma$ parameter in the $V$, $I$, and $K$ bands and in the $W_{JK}$,$W_{VK}$, and $W_{VI}$ Wesenheit magnitudes. More recently, a stronger effect
($\gamma \sim -0.2$ mag/dex) ---albeit weaker than the effect found in the present work--- was found in the same bands by \citet{Gieren2018} from a BW analysis of DCEPs in the Galaxy, LMC, and SMC, which were used to extend the range of metallicity of the pulsators adopted in the analysis. 

The advent of \gaia\ DR2 parallaxes permitted us to obtain the first reliable evaluation of the $\gamma$ term using only Galactic DCEPs with metallicities from HiRes spectroscopy. In particular, \citet{Groenewegen2018} and later \citet{Ripepi2019,Ripepi2020} found  $\gamma \sim -0.1-0.4$ mag/dex in a variety of bands and Wesenheit magnitudes (see Fig.~\ref{fig:literature_metallicity}), which is closer to the values found in this paper, but with a significance of generally lower than 1$\sigma$, owing to the still insufficient precision of DR2 parallaxes. The improved \gaia\ EDR3 parallaxes instead allowed us to obtain larger (in an absolute sense) $\gamma$ values in the first paper of the C-MetaLL project  \citep{Ripepi2021a} and for the $W_G$ magnitude \citep{Ripepi2022a}.

Other studies \citep[][]{2004A&A...420..655G,2007A&A...476...73F,2017ApJ...842..116W,Breuval2021,Breuval2022}, together with the already quoted \citet{Gieren2018}, compared the properties (i.e. the zero points of the $PL$ or $PW$ relations) of the DCEPs in the MW and in the more metal-poor galaxies LMC and SMC to estimate the extent of the $\gamma$ value. In particular, \citet{Breuval2021,Breuval2022}, used geometric distances for the DCEPs in the MW (from \gaia\ EDR3 parallaxes) and in the Magellanic Clouds (from eclipsing binaries) to estimate $\gamma$ in the same bands and Wesenheit magnitudes as those treated in the present work, after fixing the slope of the $PL$ and $PW$ relations to that of the LMC, with results ranging between $-0.178$ and $-0.462$ mag/dex.
In the context of the SH0ES project \citep{Riess2016,Riess2019,Riess2021,Riess2022a}, the metallicity effect is one of the outputs of the process for the estimation of $H_0$. In these cases, the $\gamma$ coefficient in the $W_{H, V-I}^{HST}$ magnitude used by the SH0ES team is of the order of $-0.2$ mag/dex, while we obtain a slightly larger (in an absolute sense) value in our best case, that is, with the Lit.+Gaia sample without the $\delta$ term (see e.g. Fig.~\ref{fig:coeffs-vs-wlen-abc-F1O-sflag2}). \\ 
Recent theoretical studies \citep[][]{Anderson2016,DeSomma2022} also predict mild effects of the metallicity, with $\gamma$ ranging from $-0.27$ to $-0.13$ mag/dex, although we note that the models by \citet[][]{DeSomma2022} only deal with $PW$ relations. \\
We notice that in almost all cases, we find larger $\gamma$ values (in an absolute sense) than the literature, although most of them are distributed along the lower limit in the literature of $-0.4$ mag/dex (see also Fig.~\ref{fig:coeffs-vs-wlen-abcd-F1O-sflag2}). On the other side, for the $V$-band a smaller coefficient is found, in agreement with the upper limit of $-0.2$ mag/dex. 

\section{Conclusion}

\label{sec:conclusion}

In this fourth paper of the C-MetaLL series, we studied the metallicity dependence of the $PL$ relations in the following bands: $G_{BP},G_{RP},G,I,V,J,H$, and $K_S$, and of the $PW$ relations in the following Wesenheit magnitudes: $W_{G, BP-RP},W_{H, V-I},W_{H, V-I}^{cHST}$,$W_{J, J-K}$, and $W_{V, V-K}$. 
To this end, we exploited the literature to compile a sample of 910 DCEPs with [Fe/H] measured from HiRes spectroscopy and complemented it with a number of stars with metallicity measurements based on the \gaia\ RVS instrument as released in DR3. For all these stars, we provide a table with photometry in the $G_{BP},G_{RP},G,I,V,J,H$, and $K_S$ bands, metallicity, and astrometry from \gaia\ DR3. 

We carried out our analysis adopting two different samples: one composed only of literature data and one including both the latter and \gaia\ DR3 results.
In order to estimate the parameters of the $PLZ/PWZ$ relations, we used the ABL formalism, which allowed us to treat the parallax $\varpi$ linearly and to preserve the statistical properties of its uncertainties. We considered two functional forms for the $PLZ$ and $PWZ$ relations including (i) the metallicity dependence on the intercept only (three-parameter case); and (ii) the metallicity dependence on both intercept and slope (four-parameter case). 
The main findings of our analysis can be summarised as follows: 
\begin{enumerate}

\item 
Regarding $PLZ$ relations, both $\alpha$ and $\beta$ show a linear dependence on wavelength. We provide the linear relationships between these quantities and $\lambda^{-1}$. 
These relations do not change considerably when we use the Lit.+Gaia or Lit. samples, nor when we carry out a three- or four-parameter fit. Because of the low wavelength coverage, no clear dependence can be discussed for the $PWZ$ relations.
\item 
A clear negative dependence of the intercept on metallicity ($\gamma$-coefficient) is found for all the $PL$ and $PW$ relations. For the three-parameter solutions, the values of $\gamma$ are around $-0.4:-0.5$ dex with no clear dependence on wavelength. Thus, this work confirms our previous results reported by \citet{Ripepi2021a,Molinaro2023}. For the four-parameter solutions, the values of $\gamma$ generally slightly decrease (in an absolute sense), especially for the Lit. only sample. 
In general, the $\gamma$ coefficients found in this work are larger (in an absolute sense) than those in the literature, which range between $-0.2$ and $-0.4$ dex.
\item 
The dependence of the slope on metallicity ($\delta$ coefficient) remains undetermined. Indeed, in about half of the cases, $\delta$ assumes values comparable with zero within 1$\sigma$ for the Gaia+Lit. sample,  while, especially in the Lit. case,  $\delta$  generally assumes a positive value comprised between 0 and +0.5. 

\item 
The main difference between using the F and F+1O samples is that larger error bars are found in the former case, especially for the $\gamma$ and $\delta$ coefficients.
\item 
The inclusion of global zero point offset to the individually corrected parallaxes according to L21 has a larger impact on the $\gamma$ coefficients than on the $\delta$ ones. More specifically, the adoption of offsets by $-14$ $\rm \mu$as \citep[][]{Riess2021} and $-22$ $\rm \mu$as \citep{Molinaro2023} implicates smaller and smaller values of $\gamma$ (in an absolute sense), which is in better agreement with recent literature. However, if we use the geometric distance of the LMC by \citet{Pietrzynski2019} as a reference and calculate the distance of this galaxy using our relations with and without the global offset, we find that good agreement for the distance of the LMC is found for values of the offset in between null and the 14 $\rm \mu$as values. It is worth noting that for the $W_{cH, V-I}^{cHST}$ magnitude \citet{Riess2021} offset represents the best correction. These results support larger values (in an absolute sense) of the $\gamma$ value. 
\item
We investigated the possible effect of an uneven distribution in the metallicity of the sample used in this work by resampling our data set in order to have a balanced number of DCEPs at every metallicity value. We carried out the fitting procedure on 10,000 samples extracted from the total sample, obtaining a value for the four coefficients of the fit for every data set. The medians of the obtained distributions for all the coefficients appear to be in excellent agreement with those obtained using the entire sample.
Therefore, we can conclude that the results obtained in this work are not affected by the sample's unbalanced distribution in metallicity. 
\end{enumerate}

The results presented in this paper show the importance of the extension of the metallicity range when carrying out the analysis. Further observations in order to gather HiRes spectroscopy of MW metal-poor DCEPs will offer the opportunity to better constrain the dependence on the metallicity of both the intercept and the slope. In particular, regions in the Galactic anti-centre direction are the most promising targets for future observations, where DCEPs are expected to have [Fe/H]$<-0.3:-0.4$ dex, and can therefore be used to further populate the metal-poor tail of the DCEP distribution.

%

\begin{acknowledgements}
We thank our anonymous Referee for their helpful comments which helped us to improve the paper.
We warmly thank A. Riess for useful discussion on this paper and L. Breuval for indicating us the presence of a typo in one of equation from the literature that we used in this work.    
Part of this work was supported by the German
\emph{Deut\-sche For\-schungs\-ge\-mein\-schaft, DFG\/} project
number Ts~17/2--1.

This work has made use of data from the European Space Agency (ESA) mission
Gaia (\url{https://www.cosmos.esa.int/gaia}), processed by the Gaia
Data Processing and Analysis Consortium (DPAC,
\url{https://www.cosmos.esa.int/web/gaia/dpac/consortium}). Funding for the DPAC
has been provided by national institutions, in particular, the institutions
participating in the  Gaia Multilateral Agreement.

This research has made use of the VizieR catalogue access tool, CDS, Strasbourg, France.
This research has made use of the SIMBAD database, operated at CDS, Strasbourg, France.
I.M. acknowledges partial financial support by INAF, Minigrant "Are the Ultra Long Period Cepheids cosmological standard candles?"
We acknowledge funding from INAF GO-GTO grant 2023 "C-MetaLL - Cepheid metallicity in the Leavitt law" (P.I. V. Ripepi)
\end{acknowledgements}

\bibliographystyle{aa} 
\bibliography{myBib} 

\begin{appendix}
\section{PLZ/PWZ coefficients for the F pulsators only sample}\label{appA}
For completeness, we provide figures that are analogous to Fig.~\ref{fig:coeffs-vs-wlen-abcd-F1O-sflag2}, Fig.~\ref{fig:coeffs-vs-wlen-abc-F1O-sflag2}, Fig.~\ref{fig:coeffs-vs-wlen-abcd-F1O-sflag0}, and Fig.~\ref{fig:coeffs-vs-wlen-abc-F1O-sflag0}, showing the fit results of the sample Lit.+Gaia and Lit. including only the F pulsators.

\begin{figure}
\includegraphics[width=9cm]{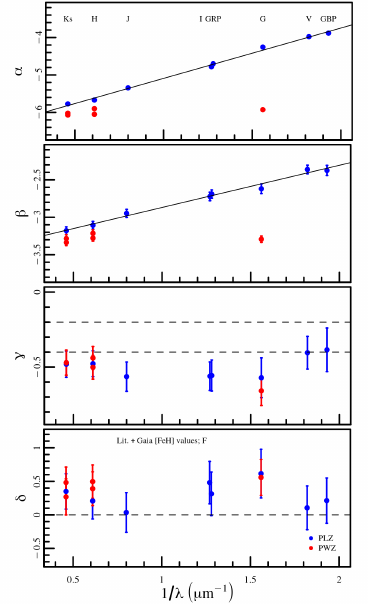}
\caption{Same as Fig.~\ref{fig:coeffs-vs-wlen-abcd-F1O-sflag2} but for the Lit.+Gaia data set including only F pulsators.}
\label{fig:coeffs_vs_wlen_abcd_F_sflag2}
\end{figure}

\begin{figure}
\includegraphics[width=9cm]{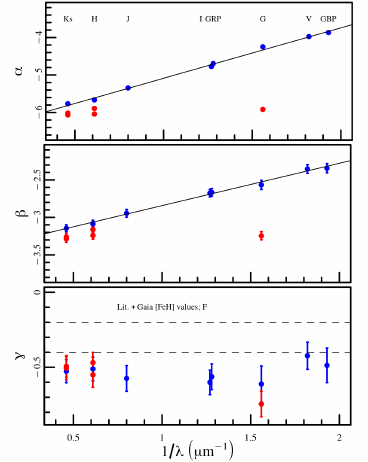}
\caption{Same as Fig.~\ref{fig:coeffs-vs-wlen-abc-F1O-sflag2} but for the Lit.+Gaia data set including only F pulsators.}
\label{fig:coeffs_vs_wlen_abc_F_sflag2}
\end{figure}

\begin{figure}
\includegraphics[width=9cm]{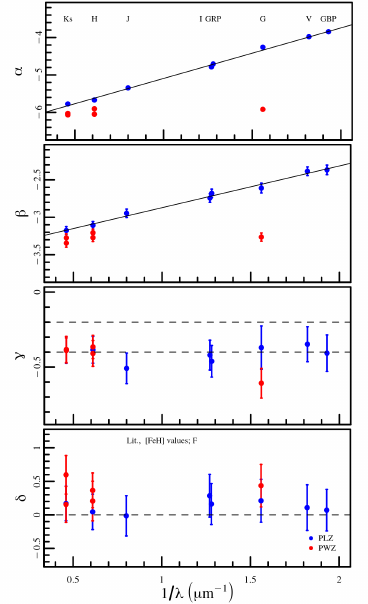}
\caption{Same as Fig.~\ref{fig:coeffs-vs-wlen-abcd-F1O-sflag0} but for the Lit. data set including only F pulsators.}
\label{fig:coeffs_vs_wlen_abcd_F_sflag0}
\end{figure}

\begin{figure}
\includegraphics[width=9cm]{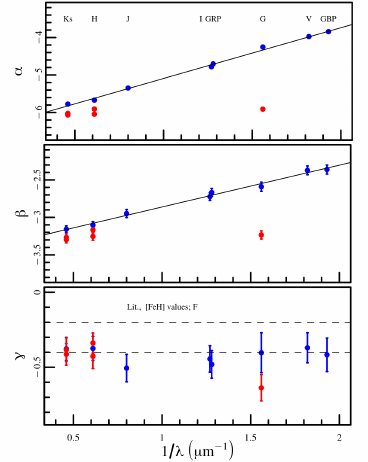}
\caption{Same as Fig.~\ref{fig:coeffs-vs-wlen-abc-F1O-sflag0} but for the Lit. data set including only F pulsators.}
\label{fig:coeffs_vs_wlen_abc_F_sflag0}
\end{figure}

\section{Parallax correction for the Lit. sample}\label{appB}
Here, we provide figures  analogues to Fig.~\ref{fig:OffsetComparisonLitGaia.pl}, Fig.~\ref{fig:OffsetComparisonLitGaia.pw}, and Fig.~\ref{fig:OffsetComparisonLitGaia.lmcDistance} for the Lit. sample.
  \begin{figure}
   \includegraphics[width=9.0cm]{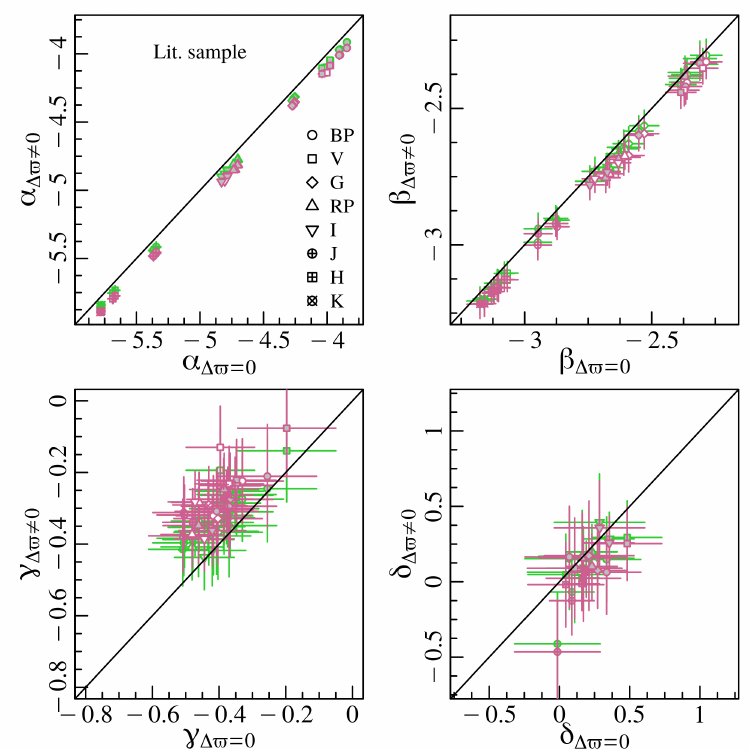}
   \caption{Same as Fig.~\ref{fig:OffsetComparisonLitGaia.pl} but for the Lit. sample.}
\label{fig:OffsetComparisonLit.pl}
   \end{figure}

    \begin{figure}
   \includegraphics[width=9.0cm]{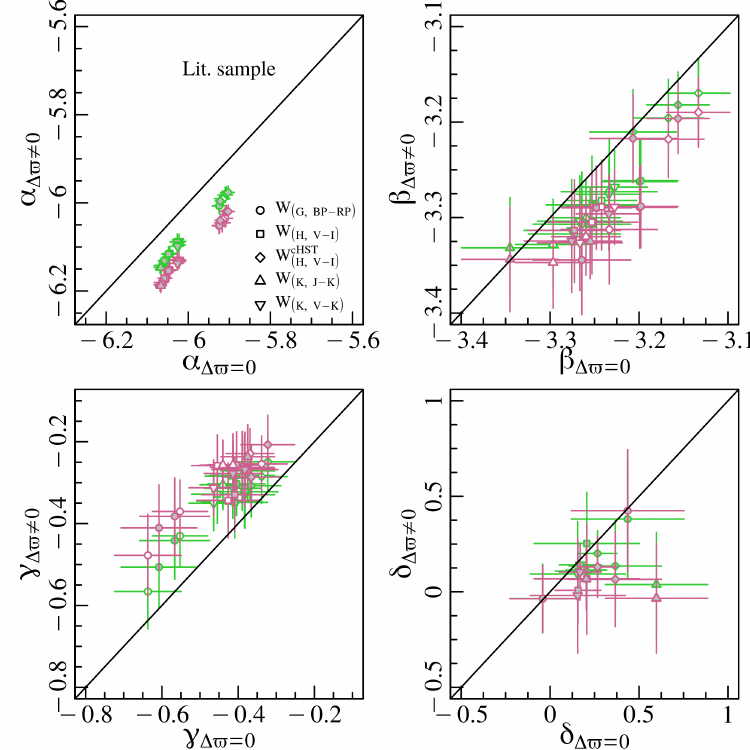}
   \caption{Same as Fig.~\ref{fig:OffsetComparisonLitGaia.pw} but for the Lit. sample.}
\label{fig:OffsetComparisonLit.pw}
   \end{figure}

   \begin{figure}
   \includegraphics[width=9.0cm]{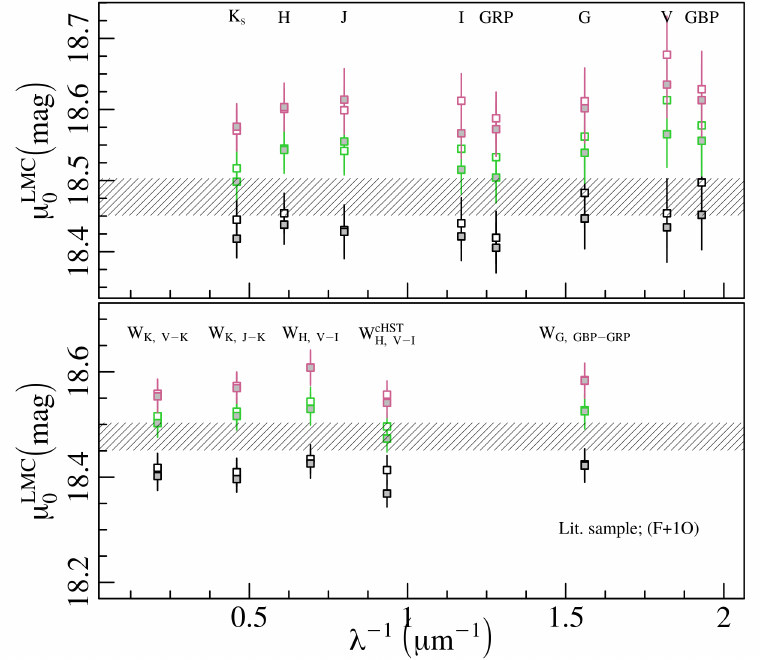}
   \caption{Same as Fig.~\ref{fig:OffsetComparisonLitGaia.lmcDistance} but for the Lit. sample.}
\label{fig:OffsetComparisonLit.lmcDistance}
   \end{figure}

\end{appendix}

%
%

\end{document}